\documentclass[preprint,12pt,3p]{elsarticle}

\usepackage{matlab-prettifier}
\usepackage{xfrac}
\usepackage{graphicx}
\usepackage{parskip}
\usepackage{hyperref}
\usepackage{longtable}
\usepackage{adjustbox}
\usepackage{setspace}
\hypersetup{ 
backref=true, 
bookmarks=true, 
unicode=false, 
pdftoolbar=true, 
pdfmenubar=true, 
pdffitwindow=false, 
pdfstartview={FitH}, 
pdftitle={}, 
pdfauthor={}, 
pdfsubject={}, 
pdfkeywords={}{}, 
pdfnewwindow=true, 
colorlinks=true, 
linkcolor=BlueViolet, 
citecolor=maroon, 
urlcolor=blue,
}
\usepackage{comment}
\usepackage{subfig}
\usepackage{caption}

\usepackage{amssymb}
\usepackage{gensymb}

\usepackage{natbib}
\biboptions{comma,round}
\setcitestyle{authoryear}

\begin{document}
\begin{frontmatter}

\title{THE ECONOMIC IMPACT OF WEATHER AND CLIMATE\footnote{Marco Letta expertly assisted with data collection and regressions. Max Auffhammer, Peter Dolton, Jurgen Doornik, Bill Greene, David Hendry, Andrew Martinez, Pierluigi Montalbano and Felix Pretis had excellent comments on earlier versions of this work. I also thank seminar participants at the EMCC-III conference, University of Sussex, UNU-MERIT University, the 2019 IAERE conference, Cambridge University, London School of Economics, Sapienza University of Rome, Shanghai Lixin University, Edinburgh University, University of Southern Denmark, Federal Reserve Bank of San Francisco, and CESifo.}}

\author[label1,label2,label3,label4,label5,label6]{Richard S.J. Tol}
\address[label1]{Department of Economics, Jubilee Building, University of Sussex, Falmer, BN1 9SL, United Kingdom; r.tol@sussex.ac.uk}
\address[label2]{Institute for Environmental Studies, Vrije Universiteit, Amsterdam, The Netherlands}
\address[label3]{Department of Spatial Economics, Vrije Universiteit, Amsterdam, The Netherlands}
\address[label4]{Tinbergen Institute, Amsterdam, The Netherlands}
\address[label5]{CESifo, Munich, Germany}
\address[label6]{Payne Institute for Earth Resources, Colorado School of Mines, Golden, CO, USA}

\begin{abstract}
I propose a new conceptual framework to disentangle the impacts of weather and climate on economic activity and growth: A stochastic frontier model with climate in the production frontier and weather shocks as a source of inefficiency. I test it on a sample of 160 countries over the period 1950-2014. Temperature and rainfall determine production possibilities in both rich and poor countries; positively in cold countries and negatively in hot ones. Weather anomalies reduce inefficiency in rich countries but increase inefficiency in poor and hot countries; and more so in countries with low weather variability. The climate effect is larger than the weather effect.


\end{abstract}

\end{frontmatter}

\newpage\section{Introduction}
\label{sc:intro}
Climate matters to the economy. Not in the way that classical thinkers such as Guan Zhong, Hippocrates or Ibn Khaldun, or modern thinkers such as \citet{Huntington1915} or \citet{Diamond1997} argue it does. Environmental determinism is inconsistent with the observations. There are thriving economies in the desert, in the tropics, and in the polar circle. There is destitution, too, in all these places. Climate is not destiny, but it does matter.

The prevailing view among economists, with some exceptions \citep{Bloom1998, Sachs2003, Olsson2005, Barrios2010}, is that climate does \emph{not} matter for economic development, only institutions do \citep{Easterly2003, Rodrik2004}. Some argue that climate and geography partly shaped institutions in the past, but have become irrelevant since \citep{Acemoglu2001, acemoglu2002reversal, Alsan2015}. Institutional determinism is just as inconsistent with the observations. The two halves of the Korean Peninsula and of the island of Hispaniola are powerful reminders of the importance of institutions, but climate matters for agriculture \citep{Mendelsohn1994, Schlenker2005}, for energy demand \citep{Mansur2008}, for tourism \citep{Lise2002}, for transport \citep{Koetse2009}, for labour productivity \citep{Kjellstrom2009, Zander2015}, and for health \citep{Sachs2002}\textemdash and thus for the economy as a whole.

Climate matters, but it has been an empirical challenge to demonstrate this using country data. Climate changes only slowly over time, its signal swamped by confounders, many of which change more quickly than climate. Climate varies substantially over space, but so do a great many other things that we know are important for development.\footnote{\citet{Druckenmiller2018} propose the solve the confounding problem by \emph{spatial} differencing, which works if the variable of interest changes at a finer resolution than its confounders, but may work less well if data are measured on different and irregular grids.} The insignificance of climate variables in cross-country studies may be due to a lack of statistical power. Indeed, a climate association is significant in \emph{subnational} income data \citep{Nordhaus2006, Dell2009, Desmet2018, Henderson2018, Kalkuhl2020, Conte2020, Alvarez2021} and, as is shown below, in long panels. Because of the confounders,\footnote{\citet{Andersen2016} argue that it is UV radiation, rather than climate, that affects development.} this association cannot be given a causal interpretation.

Unlike the impact of \emph{climate}, the impact of \emph{weather} can be identified\textemdash or so people have argued. Identification rests on the fact that weather is random \citep{Heal2016}, at least from the perspective of the economy. The problem with this argument is that by now many different economic activities have been found to be affected by the weather \citep[see][among others]{Auffhammer2011, Barreca2016, Dechenes2007, GraffZivin2020, Leightner1999, Li2018, Pechan2014, Ranson2014, Zhang2018}, and these activities impact one another.footnote{\citet{Merel2021} shows that non-linear panels with weather but no climate variables are biased.}

Causality notwithstanding, these studies show that weather matters to the economy. However, the impacts of weather shocks cannot readily be extrapolated to the impacts of climate change \citep{Dell2014, Kolstad2020}. Climate is what you expect, weather is what you get. Weather are \emph{draws} from an probability distribution. Climate \emph{is} that distribution. Climate change shifts the moments of the weather distribution \citep{auffhammer2018quantifying}. Weather is unpredictable for more than a few days ahead. Adaptation to weather shocks is therefore limited to immediate responses\textemdash put up an umbrella when it rains, close the flood gates when it pours. Adaptation to climate change extends to changes in the capital stock\textemdash buy an umbrella, build flood gates. Furthermore, adaptation to climate change depends on updates of the expectations for weather \citep{Severen2018, Lemoine2017b, Bogmans2017}. In other words, weather studies estimate the \emph{short-run} elasticity, whereas the \emph{long-run} elasticity is needed to estimate the impact of climate change.

\citet{Hsiang2016} and \citet{Deryugina2017} argue that the marginal effect of a weather shock equals the marginal climate effect. Climate change is not marginal but its total impact is an integral of marginals. Their assumptions are quite restrictive, however. Economic agents need to be (1) rational and their adaptation investments (2) optimized. Adaptation needs to be (3) private and adaptation options (4) continuous. The economy needs to be in a (5) spatial equilibrium and (6) markets complete. Adaptation investments are often long-lived, so both spot and future markets should be complete. Spatial zoning and transport hubs distort the spatial equilibrium. Adaptation is often lumpy, be it air conditioning or irrigation. Some adaptation options, such as coastal protection, are public goods. Other adaptation options, such as protection against infectious disease, have externalities. Agents are not always rational, and decisions suboptimal. The result by Deryugina and Hsiang is almost an impossibility theorem.

Weather affects economic activity, and so the measurement of the impact of climate on economic activity. Weather can be seen as noise, but that noise may well be correlated with climate, the right-hand-side variable of interest. I therefore propose a new way to simultaneously model the impact of climate and weather, to show that both matter and that previous work is misspecified.

The empirical strategy rests on the following assumptions. Climate affects production possibilities. This is obvious for agriculture: Holstein cows do well in Denmark but jasmine rice does not; the reverse is true in Thailand. Climate also affects energy and transport, and thus all other sectors of the economy. Weather affects the realization of the production potential. Hot weather may slow down workers, frost may damage crops, floods may disrupt transport and manufacturing. Conceptualized thus, climate affects the production frontier, and weather the distance from that frontier. The econometric specification is therefore a stochastic frontier analysis with weather variables in inefficiency and climate variables in the frontier.\footnote{\citet{Kumar2019} estimate the impact of temperature and rainfall on inefficiency in output \emph{growth}. I here study inefficiency in output. They omit climate from the frontier.} Climate affects potential output, weather the output gap.

I apply the proposed method to a panel of output per worker, measured at the country level. \citet{Dell2012}, \citet{Letta2018} and \citet{newell2018} find that weather shocks hit the economic growth of \textit{poorer} countries harder. \citet{Burke2015} instead find that \textit{hotter} countries are hit harder, a specification adopted by \citet{Pretis2018}, \citet{Henseler2019} and \citet{Kalkuhl2020}. \citet{Generoso2020} has a similar result. Within sample, it is difficult to distinguish between these two specifications as hotter countries tend to be poorer. However, out of sample, a hotter, richer world would be \textit{more} vulnerable to weather shocks according to Burke, but \textit{less} vulnerable according to Dell. \citet{Kahn2019} reject heterogeneity. The results below shed new light on these questions.

\citet{Moore2014} regress farm profits on the thirty-year average temperature and rainfall, and the quadratic deviation from that average, thus accounting for both climate and weather. \citet{Heutel2020} regress mortality on weather, but interact the weather effect with climate zones. \citet{Auffhammer2018} proposes a two-level hierarchical model with the impact of weather at the bottom and its interaction with climate at the top.\footnote{\citet{Bigano2006} use a similar model for tourist destination choice, with climate at destination at the bottom and climate at origin at the top.} In the model below, climate and weather interact too, but in a more intuitive way: Climate affects potential output, weather the output gap; the impact of climate is deterministic, while the effect of weather is stochastic.

The cross-validation study of \citet{newell2018} finds that weather affects the level of GDP rather than its growth rate, a specification adopted here in line with the intuition sketched above. Furthermore, I assume that the economy is affected by \emph{unusual} weather rather than weather. Frost of -10\celsius{} brought Texas to a standstill in February 2021, but is a regular occurrence in North Dakota without major consequences. I therefore standardize the weather, expressing temperature and precipitation in standard deviations from the mean. This introduces an interaction between weather and climate, and an implicit model of adaptation.

The paper proceeds as follows. Section \ref{sc:methods and data} describes methods and data. Section \ref{sc:results} presents the baseline results. Section \ref{sc:sensitivity} conducts the sensitivity analysis. Section \ref{sc:implications} discusses the implications for climate change. Section \ref{sc:conclude} concludes.

\section{Methods and data}
\label{sc:methods and data}
\subsection{Methods}
\label{sc:methods}
I assume a Cobb-Douglas production function:
\begin{equation}
\label{eq:CD}
    {Y}_{c,t} = A_{c,t} K_{c,t}^\beta L_{c,t}^{1-\beta}
\end{equation}
Total factor productivity $A_{c,t}$ is the Solow residual in country $c$ at time $t$: It captures everything that affects output $Y_{c,t}$ that cannot be explained by capital $K_{c,t}$ or labour $L_{c,t}$. 

I concentrate Equation (\ref{eq:CD}) by dividing $K$ and $L$ by labour force $L$, and denote the resulting variables in lower case.

Taking natural logarithms, the equation to be estimated is: 
\begin{equation}
\label{eq:log}
    \ln y_{c,t} = \alpha +\beta \ln k_{c,t}
\end{equation}

I assume that total factor productivity is a function of moving averages of weather variables (average temperature, $\bar T_{c,t}$, and precipitation, $\bar R_{c,t}$). This is loosely based on \citet{Nordhaus1992}. Weather shocks affect the variance of the stochastic component of permanent income. Hence, Equation (\ref{eq:log}) becomes:
\begin{equation}
\label{eq:frontier}
    \ln y_{c,t} = \beta_1 \ln k_{c,t} + f \left (\bar T_{c,t}, \bar R_{c,t} \right ) + \mu_{c} + t + v_{c,t} - u_{c,t}
\end{equation}
where $\bar T_{c,t}$ and $\bar R_{c,t}$ are the \emph{average} temperature c.q. precipitation in country $c$ in the thirty years preceding year $t$, $\mu_{c}$ is a full set of country fixed effects, \textit{t} is a linear time trend, $v_{c,t} \sim \mathcal{N}(0,\sigma^2_v)$ and
\begin{equation}
\label{eq:inefficiency}
    u_{c,t} \sim \mathcal{E} (\lambda_{c,t}) = \mathcal{E} \left (\gamma_0 + \gamma_1 g \left ( \frac{T_{c,t}-\bar T_{c,t}}{\tau_{c,t}} \right ) + \gamma_2 g \left ( \frac{R_{c,t}-\bar R_{c,t}}{\rho_{c,t}} \right ) \right )
\end{equation}
where $\tau$ and $\rho$ are the standard deviations of temperature and rainfall, respectively. Instead of the unwieldy $ \sfrac{T-\bar T}{\tau}$, I write $z(T)$; ditto for $R$. This is standardized temperature and precipitation. In the base specification, $f \left (\bar T_{c,t}, \bar R_{c,t} \right ) \equiv \beta_2 \bar T_{c,t} + \beta_3 \bar T_{c,t}^2 + \beta_4 \bar R_{c,t} + \beta_5 \bar R_{c,t}^2 + \beta_6 \bar T_{c,t} \bar R_{c,t}$, a second-order Taylor approximation, and $g(\cdot) \equiv | \cdot |$. I refer to Equation (\ref{eq:frontier}) as the frontier or potential output, and to Equation (\ref{eq:inefficiency}) as inefficiency or the output gap.

Note that in this specification, the impact of weather is stochastic. Unusual weather affects the mean and standard deviation of the output gap.

I use the True Fixed-Effect (TFE) model \citep{greene2005fixed} to estimate a one-step stochastic frontier model in a fixed-effect setting with explanatory variables in the inefficiency parameter. I use the \textsc{sfmodel} package for Stata \citep{kumbhakar2015practitioner} to estimate the model.

Equation (\ref{eq:frontier}) assumes that both error terms are stationary. This is a tall assumption.\footnote{Taking first differences of all variables may get rid of unit roots in the frontier but would change the distributional assumptions in inefficiency.} I am not aware of any statistical test for stationarity that applies to this particular estimator and these distributional assumptions.\footnote{Rob Engle (personal communication) suggests that standard stationary tests would roughly apply here.} I use three remedies. First, I include a time trend in Equation (\ref{eq:frontier}), and try many variants of that trend. Second, I show robustness to different specifications. Third, I reformulate the model as an error-correction one. The output gap follows
\begin{equation}
\label{eq:errcorr}
    \Delta \ln y_{c,t} = \psi_1 \Delta z \left ( T_{c,t} \right ) + \psi_2 \Delta z\left ( R_{c,t} \right ) + \psi_3 V_{c,t} + \mu_{c} + w_{c,t} 
\end{equation}
where potential output is
\begin{equation}
\label{eq:coint}
    V_{c,t} =  \ln y_{c,t} - \mu_{c} - \mu_{t} - \vartheta_1 \ln k_{c,t} - f( \bar T_{c,t},  \bar R_{c,t} )
\end{equation}
and $\mu_{t}$ are time dummies which act as a non-parametric time trend.\footnote{The use of a non-parametric time trend was not possible in the baseline SFA model because the inclusion of so many time dummies causes convergence issues in an already computationally cumbersome maximum likelihood estimation.} This alternative estimation strategy shows that the findings are robust to the inclusion of non-parametric time trends. This alternative specification is also better suited to explicitly model the path of convergence towards the long-term equilibrium in a stochastic setting and provide empirical evidence for the speed of recovery after weather perturbations. I of course also perform the usual stationarity tests on the error-correction model.

I test for heterogeneity by interacting the variables of interest with dummies for poor countries and hot countries. I define a country as ``poor" if the World Bank does.\footnote{The WB classification of high-income economies is available \href{https://datahelpdesk.worldbank.org/knowledgebase/articles/906519-world-bank-country-and-lending-groups}{here}.} Alternative, a country is deemed poor if its GDP per capita was below the 25th percentile of the distribution in the year 1990.\footnote{1990 is the first year for which we have complete data on PPP GDP per capita for all countries. I choose the 25th percentile of the income distribution because, after testing the 25th, 50th and 75th percentiles, the specification using the 25th percentile resulted the best one according to the Wald Test.} A ``hot" country is defined as a country whose average annual temperature is above the 75th percentile of the distribution.

\subsection{Data}
\label{sc:data}
The dataset is an unbalanced panel consisting of 160 countries over the period 1950-2014.\footnote{Data are not missing randomly \citep{Baltagi2006}. Warmer countries tend to have shorter records. This is confirmed by a panel fixed-effect logit regression of having GDP data on the same climate variables as in the base specification below. The correlation between the residuals of the base specification and the logit model is -2.2\%. Selection bias is therefore minimal.} Data for this study come from two sources. Economic data on output, capital and labour force are taken from the Penn World Table (PWT), PWT 9.0 \citep{feenstra2015next}. Weather data are from the University of Delaware's \textit{Terrestrial air temperature and precipitation: 1900-2014 gridded time series, (V 4.01)} \citep{matsuura2015terrestrial}. These gridded data have a resolution of $0.5 \times 0.5$ degrees, corresponding roughly to $55 \times 55$ kilometers at the equator. Following previous literature \citep{Dell2014, Burke2015, auffhammer2013using}, we aggregate these grid cells at the country-year level, weighting them by population density in the year 2000 using population data from Version 4 of the \textit{Gridded Population of the World}.\footnote{Available \href{http://sedac.ciesin.columbia.edu/data/collection/gpw-v4}{here}.}, with the exception of Singapore.\footnote{Singapore has a surface smaller than the size of the weather grids. Given it is one of the few countries that are both rich and hot and thus increase the statistical power of the analysis, we kept it in the sample by attributing to it the weather data of the grid cell in which it is situated.} We use these weather data to construct both the climate and weather variables as defined in Section \ref{sc:methods}. Table \ref{tab:summ} presents descriptive statistics for the key variables.\footnote{See the Appendix for a complete list of countries and regions in the sample.}

\section{Results}
\label{sc:results}
Table \ref{tab:base} shows the results of the base specification outlined in Equations (\ref{eq:frontier}) and (\ref{eq:inefficiency}). Six variants are presented. Column 1 reports homogeneous effects in both the frontier and the inefficiency. In the frontier, capital per worker has a significant impact on output per worker. The output elasticity is around 0.63, in line with previous estimates. This estimate is robust to specification. Long-run temperature (i.e. climate) has a significant impact on the production frontier, but precipitation does not, as in earlier papers \citep{Dell2012, Burke2015, Letta2018}. Short-term weather anomalies, either temperature or precipitation, are insignificant in determining inefficiency.

Columns 2 and 3 show heterogeneous impacts between rich and poor countries. The hypothesis is that poor countries are disproportionately affected by climate and weather, as economic activity is concentrated in agriculture and public investment in protective measures is limited. Column 2 allows heterogeneity only in the production frontier. That is, I interact climate variables with the poor country dummy defined in Subsection \ref{sc:methods}. The interaction terms are individually insignificant. Column 3 adds heterogeneity in inefficiency. Results for the production frontier are almost unchanged. Impacts on inefficiency sharply differ among rich and poor countries: the latter suffer from large and strongly significant effect of temperature and rainfall anomalies, whereas the impact is smaller and positive in rich countries.

Column 4 adds more heterogeneity in inefficiency by interacting weather anomalies with the 'hot country' dummy defined in Section \ref{sc:methods}. These interactions are significant, and strengthen the significance of other parameters in the efficiency. Previous findings had \emph{either} poor countries \citep[e.g.][]{Dell2012, Letta2018} or hot countries \citep{Burke2015} particularly vulnerable to weather anomalies. I find \emph{both}.

Columns 1-4 specify that, in the frontier, hot and cold countries respond differently to temperature, and dry and wet countries differently to rainfall. Column 5 adds the interaction between rainfall and temperature to the frontier. This interaction is negative, but less so in poor countries. The rainfall terms are now significant too: Wetter countries are richer, and this effect is weaker for poor countries.

Dropping the insignificant interaction terms between temperature and poverty (column 6) hardly affects the parameter estimates. Column 6 is the preferred specification.

Weather anomalies increase inefficiency in poor countries, as expected. Weather anomalies \textit{decrease} inefficiency in rich countries\textemdash that is, unexpectedly much or little water, or unusually hot or cold weather stimulate the economy. This is harder to explain. It may reflect the restoration effort after floods, and crop insurance and government support after droughts. The data are GDP rather than NDP, and thus suffer from Bastiat's broken window. This effect is not observed in poor countries because restoration after natural disasters is limited and delayed \citep{Cavallo2011}.

I interpret the effect size below, after discussing the robustness of the results.

\section{Robustness}
\label{sc:sensitivity}
I implement three different types of robustness checks: sensitivity to different specifications in the SFA model; an alternative distributional assumption for the inefficiency parameter; and an error-correction model to formally test for non-stationarity. For all these sensitivity tests, with the exception of the error-correction model, I only report estimates of the preferred specification, column 6 of Table \ref{tab:base}.

\subsection{Alternative specifications}
This first set of robustness checks implements the same baseline model described in Equations (\ref{eq:frontier}) and (\ref{eq:inefficiency}) but adopts a broad set of different specification choices for key variables and interactions.

\subsubsection{Poor v rich}
I test whether the core findings are driven by the somewhat arbitrary discrimination between rich and poor countries. I replace the World Bank classification of countries that are rich\footnote{Available \href{https://datahelpdesk.worldbank.org/knowledgebase/articles/906519\#High_income}{here}.} by the  ``poorest 25\% in 1990''. Results are in column 2 of Table \ref{tab:robust}. Column 1 repeats the base specification (column 6) of Table \ref{tab:base}.

For the production frontier, results are qualitatively the same as in Table \ref{tab:base}. The main difference is that precipitation loses much of its predictive power, highlighting that different economies do respond differently to the availability of water resources. As for the inefficiency, results are again qualitatively similar to the baseline model, but coefficients are closer to zero and less significant. The log-pseudolikelihood is much lower.

\subsubsection{Squared anomalies}
Second, I replace \textit{absolute} weather anomalies in the inefficiency term with \textit{squared} anomalies. This places a heavier weight on larger anomalies. See column 3 of Table \ref{tab:robust}. The results for the production frontier are largely unaffected, and the qualitative results for the inefficiency are as above. The log-pseudolikelihood falls.

\subsubsection{Linear anomalies}
The weather anomalies in Equation (\ref{eq:inefficiency}) are absolute anomalies. Cold and hot weather, wet and dry spells are assumed to equally increase technical inefficiency. Column 4 of Table \ref{tab:robust} instead use the anomalies. Estimates for the production frontier are almost unaltered. The parameters for inefficiency become insignificant. Economies are affected by unusual weather, rather than by the weather \textit{per se}. Adaptation matters.

\subsubsection{Asymmetric anomalies}
I also test for asymmetric anomalies, disentangling negative and positive weather shocks on inefficiency. This is the preferred specification of \citet{Kahn2019}. Results are in column 5 of Table \ref{tab:robust}. The frontier is not affected. The results are much as above, with anomalous weather being good for rich countries but bad for hot and poor countries. While there is some evidence for asymmetry between the impact of wet and dry spells, cold and hot spells, the increase in the log-pseudolikelihood is minimal (less than 6 points) for the six additional parameters estimated. 

\subsubsection{Weather in the frontier}
I also look at weather effects on productivity, moving weather anomalies from the inefficiency parameter to the production frontier. Results are in column 6 of Table \ref{tab:robust}. The frontier does not change. Coefficients of weather variables are individually insignificant and the log-pseudolikelihood is sharply lower. This specification, variations of which are often used in literature, is not the preferred one.

\subsubsection{Half-normal distribution}
Equation (\ref{eq:inefficiency}) assumes an exponential half-normal distribution for inefficiency. Column 7 of Table \ref{tab:robust} show results for the half-normal distribution.\footnote{Truncated-normal models with fixed-effects are known to suffer severe convergence issues, and this case was no exception. It is therefore excluded.} The estimates for the frontier are as above. The inefficiency parameters are much the same, but the interactions with heat lose significance.

The log-pseudolikelihood falls. One key difference is that the standard deviation of the inefficiency equals its expected value for the exponential distribution, but its expected value times $\sqrt{0.5 \pi - 1}$ for the half-normal distribution. The data are overdispersed for the half-normal.

\subsection{Institutions}

\subsubsection{Capital as a substitute for climate}
I find a significant association between climate and economic performance. In the concentrated Cobb-Douglas production function, Equation (\ref{eq:CD}), there are two determinants of output per worker: climate and capital per worker. In this specification, capital is a \textit{de facto} substitute for climate, with a constant elasticity. I test that assumption, answering the question whether sufficient capital would make a country immune from the influence of its climate. I therefore interact long-run temperature variables with capital per worker in the production frontier. See Table \ref{tab:institute}, Columns 2 and 3; column 1 reproduces the base model from Table \ref{tab:base}. Rainfall is significant and so are its interactions with capital. The interactions have the opposite signs. That is, climate's influence on output shrinks as capital deepens. The interaction between temperature, rainfall and capital is insignificant. The log-pseudolikelihood increases by 7 points. However, interactions work both ways. The output elasticity of capital now depends on rainfall, varying between 0.73 in the driest countries and 0.93 in the wettest ones. A 5.5\% increase in rainfall, well within the climate change projections for this century, would lead to increasing returns to scale and explosive economic growth. I therefore keep the base specification as is.

Column 2 only changes the frontier. In column 3, I replace the interaction with the poverty dummy by an interaction with capital per worker. Signs change and the log-pseudolikelihood falls. Poverty is more than a lack of capital, and poverty drives vulnerability to weather shocks.

\subsubsection{Institutions \textit{vs} climate}
In the debate on the long-run determinants of growth and development, some find that climate plays a fundamental role in shaping long-run development, whereas others argue that the impact of climate disappears when accounting for institutions, although climate may have shaped those institutions. I test this in column 4 of Table \ref{tab:institute}. As a proxy for institutional quality, I use the \emph{Polity2 Score}.\footnote{The Polity Project Database, annual national data for the period 1800-2017, can be downloaded \href{http://www.systemicpeace.org/inscrdata.html}{here}.} This categorical variable is an aggregate score which ranges from -10 (hereditary monarchy) to 10 (consolidated democracy). While this is not the best indicator for institutional quality, it is correlated with other indicators. Historical depth is the key advantage of Polity2 over other indicators, which are available only for recent years. I interact it with long-run precipitation in the production frontier.

The results for inefficiency are essentially the same as in the base specification. In the frontier, the impact of temperature and capital is unchanged. However, the effect of rainfall is very different. Polity2 and its interactions have an insignificant effect. 

\subsection{Cointegration}
Non-stationarity is a key concern in any long panel of economic data. The residuals of the stochastic frontier model do not pass a stationarity test. See Table \ref{tab:stat}. Panel stationarity tests require that the residuals of every country are stationary. Equation (\ref{eq:frontier}) has a common trend for all countries. The panel is unbalanced, with fewer observations for hotter and poorer countries in the early years. It should therefore not come as a surprise that the model fails the test for panel cointegration.

Table \ref{tab:trend} shows the results if the model is estimated without a trend, a linear trend (as above), and a polynomial trend of order two or three; and if a different linear trend is used for poor and for other countries. Qualitatively, the impact of climate and weather is the same. The differences between estimates are not significant. Although the residuals of the alternative models are not stationary (results not shown), the stability of the results suggest that the regression results are not spurious.

Table \ref{tab:stat} supports that suggestion. Output and capital per capita are non-stationary, but the climate and weather variables are. That means that the residuals of the model are non-stationary because output and capital do not cointegrate (after inclusion of a trend). The impact of climate and weather on output per worker is not spurious\textemdash climate and weather do not explain the residual trend in output because there is no trend in the climate and weather data. 

The rightmost columns of Table \ref{tab:stat} re-estimate the model in first differences. Note that the difference between two exponential distribution is not an exponential distribution; an stochastic frontier model in first differences is a different specification. The second-to-rightmost column estimates the frontier in first-differences and adds lagged variables to inefficiency. Reassuringly, the output elasticity of capital does not significantly change when the model is estimated in first differences. The impact of weather and climate either becomes insignificant or much smaller.

In the rightmost column, I estimate the frontier in first differences, adding the first difference of the estimated inefficiencies in the base specification (column (6) in Table \ref{tab:base}, colum ``linear'' in Table \ref{tab:stat}). Inefficiency enters without explanatory variables. This is a different specification than the base one\textemdash the sum of exponential distributions is not exponential\textemdash and two-stage estimation is inefficient. That said, the signs and significance of the coefficients are as in the base specification. Estimated values are different from the base specification for temperature, precipitation and their interaction. The bottom row of Table \ref{tab:stat} shows that differencing does not solve the cointegration problems\textemdash economic growth is too variable over time and space to be captured by a simple model.\footnote{I re-estimated the model with country fixed-effects in first differences (results not shown); the impact of climate on the frontier is not materially affected, but the residuals do not become stationary.} Qualitatively, however, the results remain\textemdash the impact of climate on the frontier is not spurious.

\subsection{Error-correction model}
As a further empirical test, I estimate the error-correction model (ECM) defined in Equations (\ref{eq:errcorr}) and (\ref{eq:coint}). I assume that weather anomalies cause short-term deviations from the long-run equilibrium, while climate affects the long-run equilibrium growth path of the economy. The error-correction model is dynamic, unlike the stochastic frontier models above, tracking the time needed to absorb the perturbation caused by weather anomalies. The ECM specification allows for country and year fixed-effects, replacing the linear time trend in the stochastic frontier. 

Table \ref{tab:cv} presents the results for the long-run co-integrating vector, Table \ref{tab:ec} for the short-run error-correction. In the short-run error-correction estimates, $V$ is the residual of Table \ref{tab:ec}, Column 4, since this specification fits the data best.

The output elasticity of capital in the co-integrating vector is much the same as above. The climate variables and their interactions with the poverty dummy are not individually significant, with a few occasional exceptions, but the log-pseudolikelihood reveals that they are jointly significant: 162 points gain for 10 parameters. This is confirmed by Table \ref{tab:statecm}: Without the climate variables, the \citet{IPS2003} test firmly rejects the null-hypothesis that the residuals are stationary.

The cointegrating vector and the stochastic frontier model have the same signs on the climate variables and on their interactions with poverty. Qualitatively, the above findings are confirmed. 

Table \ref{tab:statecm} shows that the residuals of the short-run equation are stationary. Table \ref{tab:ec} shows the estimates. The cointegrating vector is highly significant. The parameter estimate of 0.06 indicates rather fast convergence to the equilibrium relationship. Precipitation is not significant but temperature is, in poor countries. This result is qualitatively different from the stochastic frontier model\textemdash but similar to \citet{Dell2012}.

Note that the results in Table \ref{tab:ec} are for the standardized temperature and precipitation, rather than their absolute values. This is a further deviation from the stochastic frontier model. Table \ref{tab:ecabs} shows the results for the absolute anomalies. The results are much the same, except that temperature now also affects rich countries. The log-pseudolikelihood is lower, however.

\section{Implications}
\label{sc:implications}
The impact of climate change is highly nonlinear in this model. The effect size is therefore hard to grasp. Furthermore, there are 160 countries in the database. There are many scenarios and models of climate change, and many scenarios and models of future economic growth. Exploring all possible futures is a combinatorial explosion, and would shed little light on how the model presented here works. So instead, I used stylized scenarios to illustrate the impact of climate change, according to column 6 in Table \ref{tab:base}, on the 2014 population, economy and climate.

The production frontier, Equation (\ref{eq:frontier}), depends on the thirty-year average of the \textit{level} of temperature and precipitation. This is projected to change over time. Inefficiency, Equation (\ref{eq:inefficiency}), depends on the absolute value of the \textit{standardized} temperature and rainfall. Without climate change, there are weather shocks to inefficiency and hence economic output. With climate change, weather shocks are different.

I consider warming between 1\celsius{} and 6\celsius, and 0.01\celsius/year and 0.06\celsius/year. This is the range shown in the Fifth Assessment Report of the Intergovernmental Panel on Climate Change (IPCC). I let rainfall increase or decrease by up to 30\%, again within the range of expectations for this century. The impact of these scenarios on the frontier is immediate. 

The impact of climate change on inefficiency follows from the deviation of the actual weather from the expected weather. Without climate change, the expected temperature shock is zero. With a 3\celsius{} per century warming, the expected temperature shock is $15 \times 0.03 / \tau_c$ per year, where the factor 15 is there because I use the 30-year average and standard deviation for normalization.

Climate affects production possibilities, and anomalous weather the realisation of those possibilities. Climate change will affect both. Extrapolating statistical models is always tricky. Here, the frontier is estimated on a wide range of climates, while inefficiency depends on time-varying standardization of weather variables. Both help to make extrapolation more reliable.

\begin{figure}[hp]
    \centering
        \includegraphics[width=0.49\textwidth]{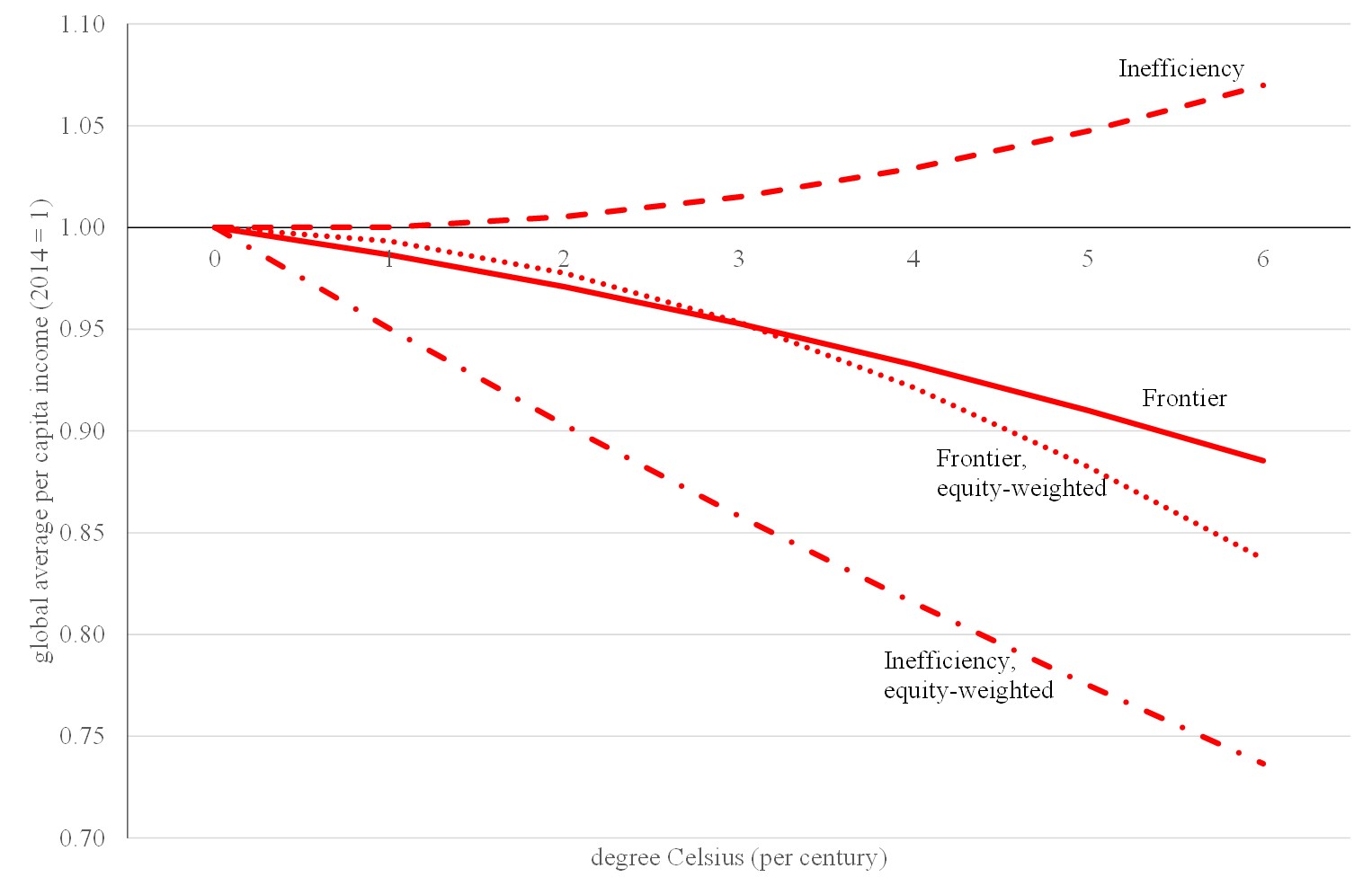}
        \includegraphics[width=0.49\textwidth]{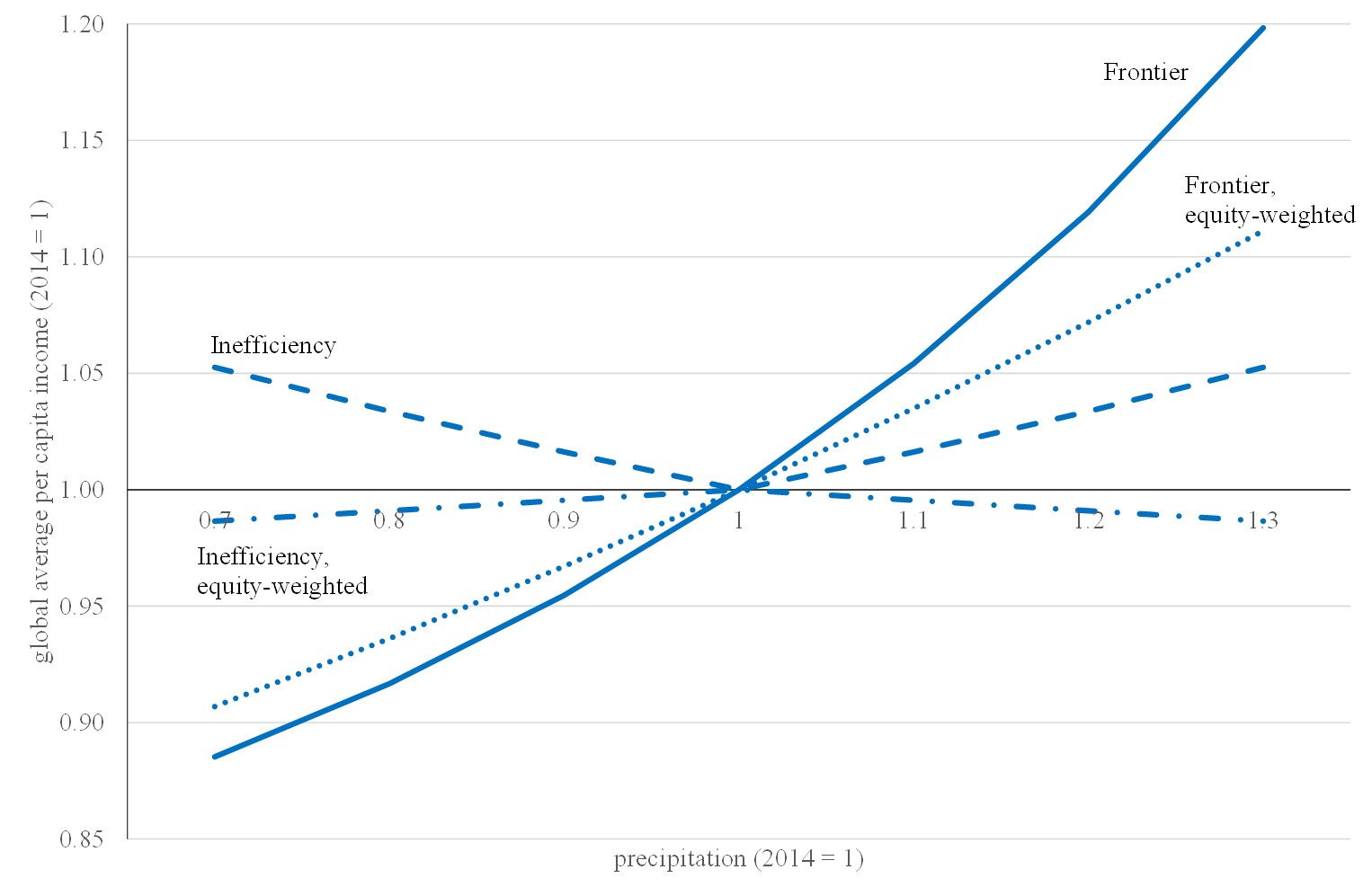}
        \caption{The change in global average output per worker due to changing temperature (left panel) and precipitation (right panel).}
    \label{fig:impact}
\end{figure}

Figure \ref{fig:impact} shows the global average impact, separately for changes in temperature and precipitation. The impact on the frontier is not out of line with previous studies \citep{Tol2018}: A 5\% loss for 3\celsius{} warming. The function is almost linear. The impact on inefficiency is more non-linear, but smaller and \emph{positive} because the impact on rich countries dominates.

This is confirmed by the second set of graphs in Figure \ref{fig:impact}. The above results compute the global average output. The two remaining graphs compute the global average utility, expressed in its income equivalent, assuming a rate of risk aversion of one \citep{Fankhauser1997}. At the frontier, these equity-weighted impact are more linear and larger if warming exceeds 3\celsius. This is because poorer countries are hit harder by climate change at the frontier. This is more pronounced in inefficiency: The sign flips, and the global average impact is substantially larger than on the frontier.

The right panel of Figure \ref{fig:impact} show the impact of changes in precipitation. At the frontier, the impacts are large. Drying would be a loss, wettening a gain. These impacts are less pronounced if the national impacts are equity-weighted. This follows from Table \ref{tab:base}: Poor, hot countries have smaller parameters. For inefficiency, change matters rather than the direction of change; inefficiency is determined by deviations from experience, regardless of whether that deviation is more or less water than expected. The impacts are more modest. Equity-weighting again flips the sign: Poor countries are negatively affected, rich countries positively.

\begin{figure}[hp]
    \centering
        \includegraphics[width=0.49\textwidth]{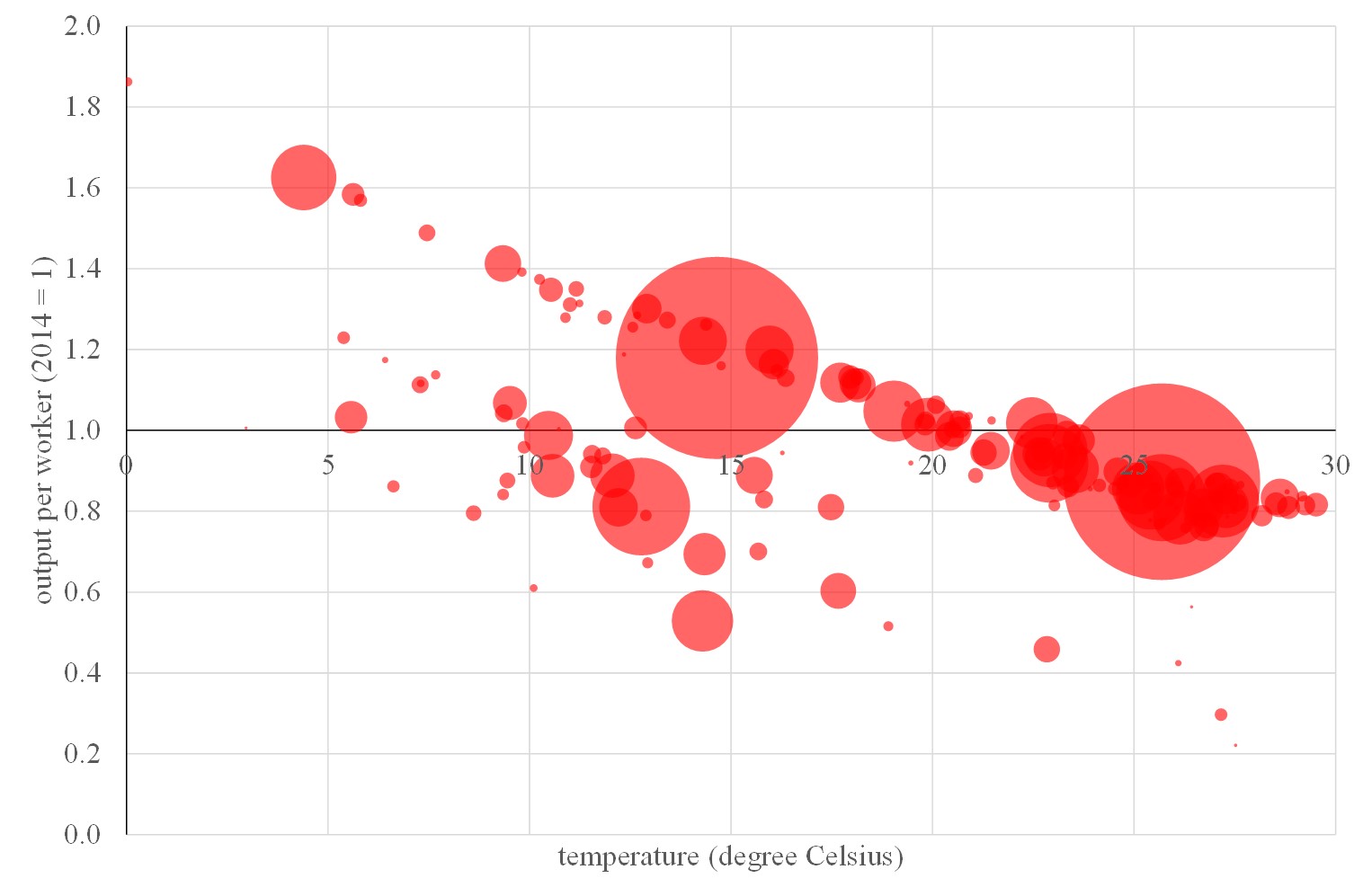}
        \includegraphics[width=0.49\textwidth]{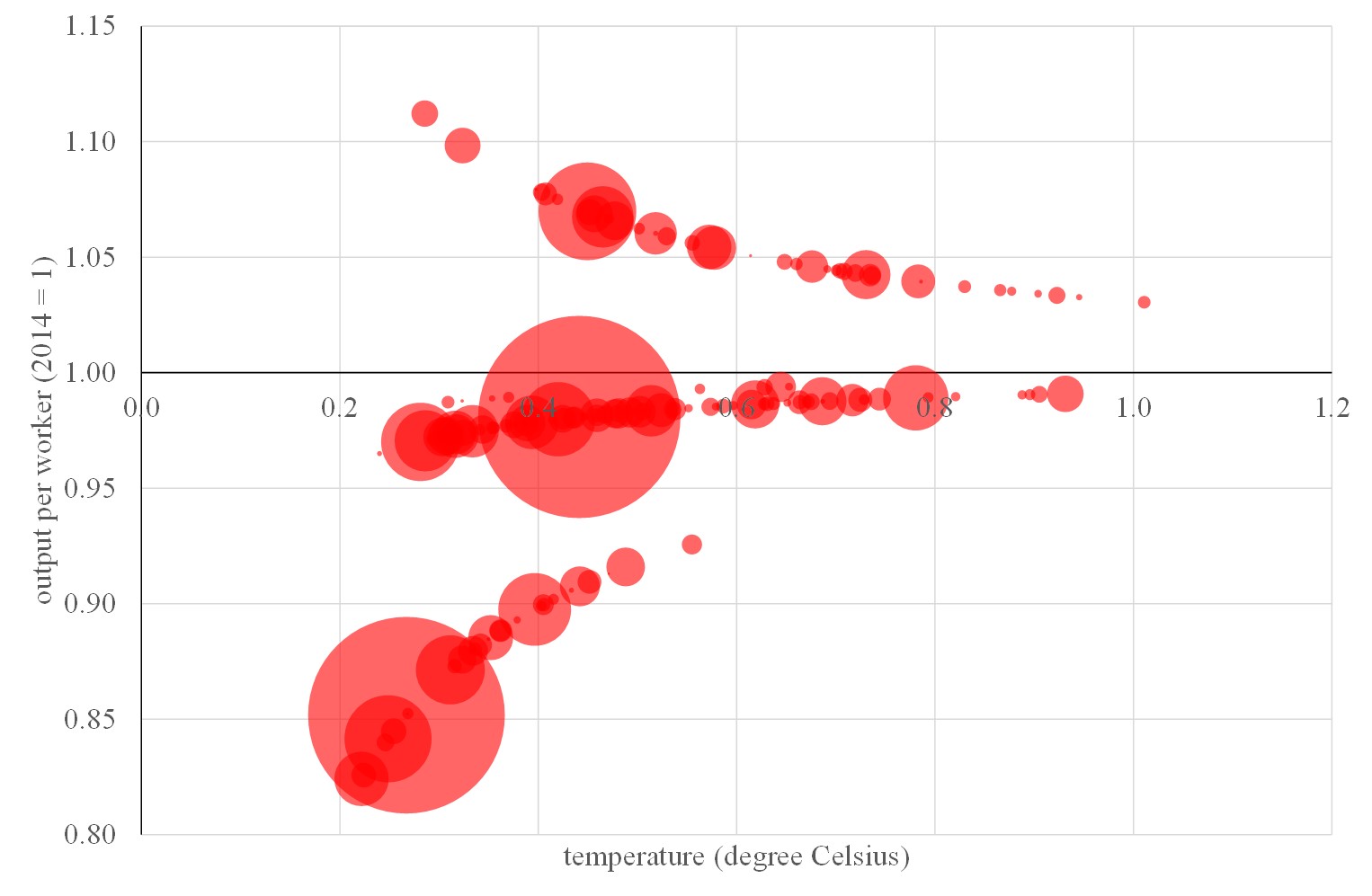}
        \includegraphics[width=0.49\textwidth]{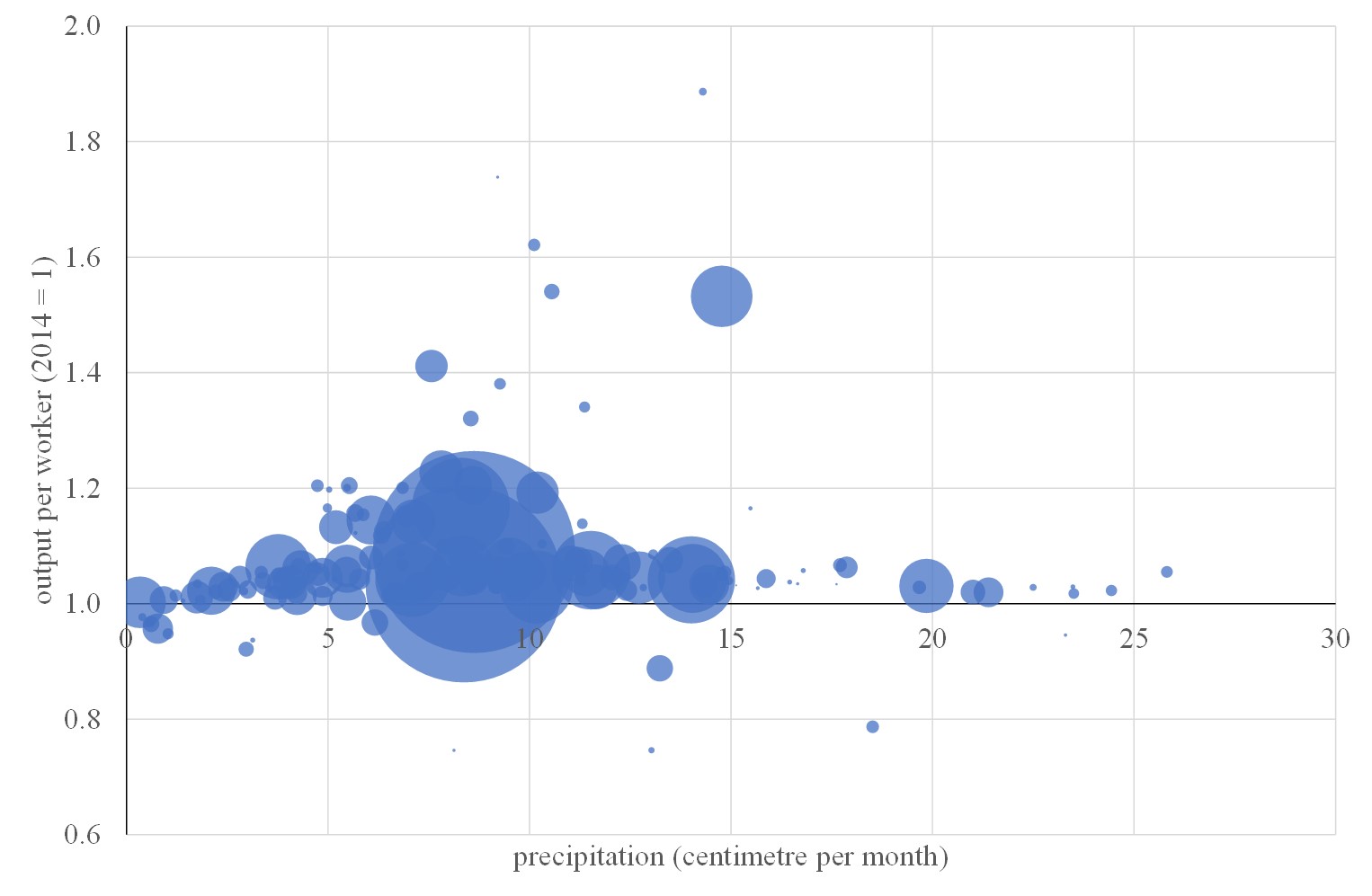}
        \includegraphics[width=0.49\textwidth]{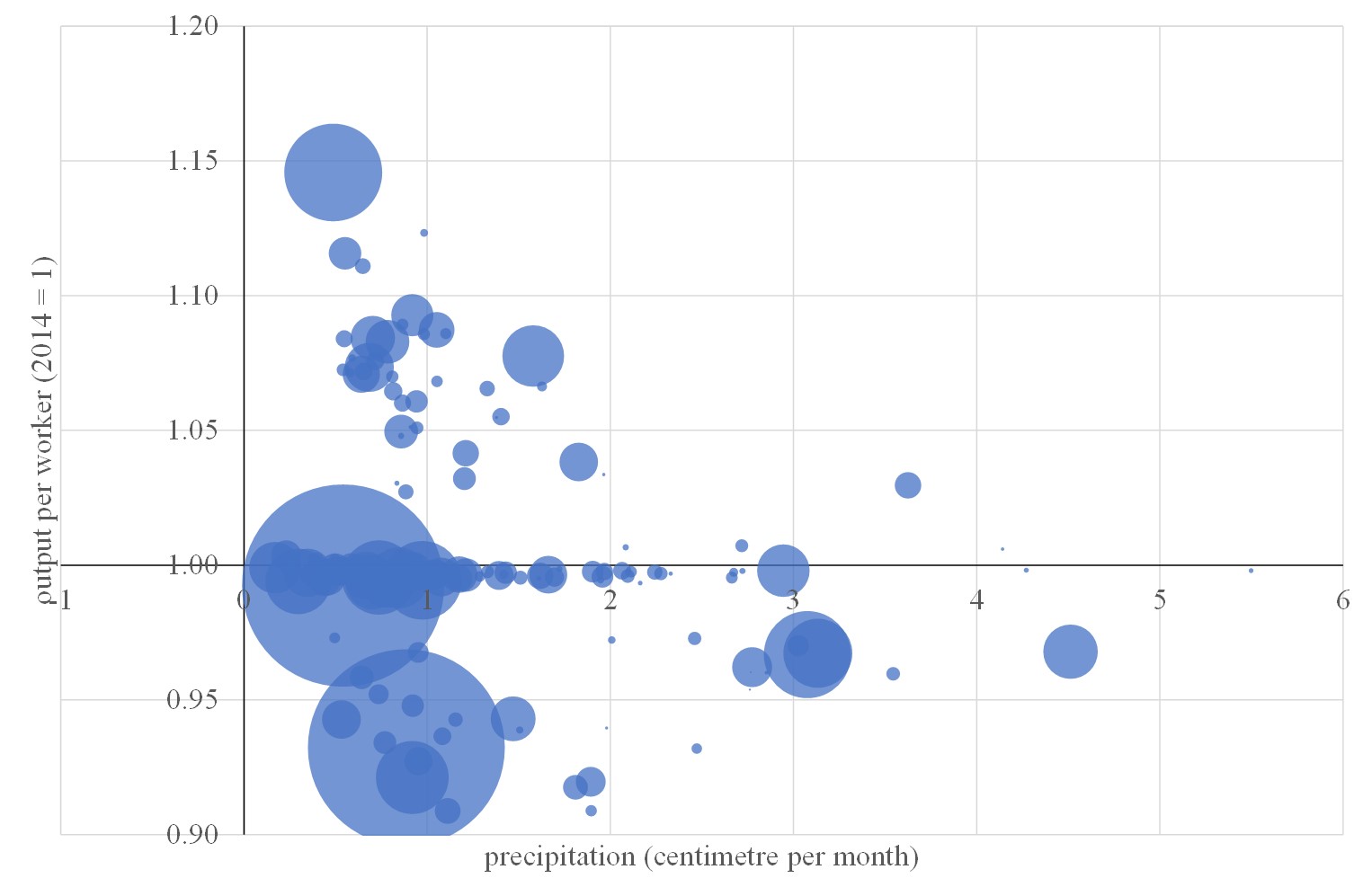}
        \caption{The change in national average output per worker due to changing temperature (top panels) and precipitation (bottom panels), in the frontier (left panels) and inefficiency (right panels). The bubble size is proportional to population size.}
    \label{fig:natimp}
\end{figure}

Figure \ref{fig:natimp} shows the results by country, for a 3\celsius{} warming and a 20\% increase in precipitation over a century. In all figures, the size of the bubble is proportional to the population size in 2014.

The top left figure shows the impact of warming on the frontier, plotted against the average temperature for 1985-2014. The spread is quite large, ranging from a 90\% increase to a 70\% decrease. Colder countries see more positive impacts, hotter countries more negative ones. The figure separates poor countries\textemdash which are essentially on a continuous lines\textemdash and rich ones\textemdash which are more dispersed because the impact of wealth is interacted with precipitation. Richer countries face more negative impacts.

The top right figure shows the impact of warming on inefficiency, plotted against the standard deviation of the temperature for 1985-2014. Effect sizes are smaller than on the frontier, ranging between a 20\% decline and a 15\% increase, and fall for countries with greater climate variability. There are three separate graphs, corresponding with the interactions in Table \ref{tab:base}. Rich countries see benefits, poor but cool countries moderate losses, and poor and hot countries large losses.

The bottom left figure plots the impact of wettening on the frontier against average precipitation in 1985-2014. Heterogeneity is again large, ranging from the 15\% loss to a 90\% gain. There is little structure in the graph.

The bottom right figure plots the impact of wettening on inefficiency against the standard deviation of precipitation in 1985-2014. Effect sizes are smaller than on the frontier, ranging between a 10\% loss and the 15\% gain, and fall with greater climate variability. There are again three separate graphs. Rich countries see gains, poor and cool countries small losses, and poor and hot countries large losses.

\section{Discussion and conclusion}
\label{sc:conclude}
I use stochastic frontier analysis to jointly model the impacts of weather and climate on economic activity in most countries over 65 years. I distinguish production potential, affected by climate, and the realisation of economic output, affected by weather. Weather shocks thus have a transient effect, climate change a permanent impact. Warming affects production potential, positively in cold, negatively in hot countries; and more so in rich, wet countries. Changes in precipitation also affect the frontier. The impacts are heterogeneous without an obvious pattern. Climate change also affects inefficiency, particularly in countries with little climate variability, reducing the output gap in rich countries but increasing it in poor and hot countries. The weather effect is small compared to the climate effect. These results are qualitatively and quantitatively robust to alternative specifications, controls, and estimators. 

\citet{Dell2012} find that poor countries are particularly vulnerable to weather shocks, \citet{Burke2015} find that hot countries are. In the Burke (Dell) specification, countries would grow more (less) vulnerable to unusual weather in a hotter and richer future. I find that both are true, and that the impact of heat is about as strong as the impact of poverty. Reduced outdoor work and manual labour, decreased relative importance agriculture in output and work force, and greater diffusion of adaptive capital such as air conditioning would help poorer countries to dampen the negative effects of weather shocks\textemdash but only to a degree, as the effort needed to alleviate the heat rises with the temperature.

The impact of weather shocks found here cannot directly be compared to previous studies. \citet{Letta2018} model economic growth as a function of the change in temperature, \citet{Dell2012, Burke2015, Pretis2018} and \citet{Kalkuhl2020} as a function of the temperature level. \citet{Kahn2019} come closest to my specification, but they use (asymmetric) weather anomalies rather than standardized weather. Another key difference with those papers is that, here, the impact of a weather shock is transitory. Unusual weather increases inefficiency, but the economy bounces back the next year, registering higher growth. If my specification is right, then previous studies that excluded lagged temperature effects are wrong.\footnote{The lags in \citet{Dell2012} are insignificant.}

Previous studies, \citet{Barrios2010} and \citet{Generoso2020} excepted, did not find a significant impact of precipitation. This is a puzzling result, as droughts and floods are more devastating than heat and cold. The same result is found here, in the frontier, unless I interact precipitation with temperature and poverty. Net water\textemdash rainfall minus evaporation\textemdash matters rather than gross water\textemdash rainfall\textemdash and more so in countries that depend more on agriculture. Precipitation also has a significant effect on inefficiency, one that varies strongly with its variability. Previous studies did not standardize weather variables.

The impact on the frontier is larger than in previous studies of the impact of climate change \citep{Tol2018}. Compared to some previous empirical studies \citep{Easterly2003, Rodrik2004}, climate has a significant effect, also when controlling for institutional quality, perhaps because I used more data \citep[as did][]{Nordhaus2006, Dell2009, Henderson2018, Kalkuhl2020}, perhaps because I \emph{modelled} heteroskedasticity. Previous studies did not do this and therefore their estimators would be inefficient and, if weather-related heteroskedasticity correlates with climate, may be biased.

Higher income, more capital nor better institutions fully insulate countries from the influence of their climate. This contradicts earlier studies \citep{Acemoglu2001, acemoglu2002reversal, Alsan2015}.

Besides the methodological advance and the new insights, the model proposed here also provides a way forward for stochastic integrated assessment models, some of which \citep[e.g.][]{Cai2019, Hambel2021} combine a \emph{deterministic} climate change impact function with \emph{stochastic} weather realisations.\footnote{See \citet{Estrada2015} for a discussion of the pitfalls and an alternative.} The framework in this paper separates the deterministic from the stochastic.

I do not include all impacts of climate change. I omit direct impacts on human welfare, such as biodiversity and health. The model does not capture the range of events which could be triggered by climate change but lie outside the current range of historical experience, such as thawing permafrost\citep{Wirths2018}, a thermohaline circulation shutdown \citep{Anthoff2016} or unprecedented sea level rise \citep{Nordhaus2019}. Because of data availability, I use democracy as a proxy for high-quality government. I limit the attention to aggregate economic activity. Adaptation and expectations are implicit in the model, as are production risks and risk preferences. The projections with respect to climate change are static, not dynamic.

The econometrics also need improvement. While cointegration does not seem to be an issue, the stationarity tests used here were not designed for the error structure assumed. I ignored heterogeneity, time-varying parameters, cross-sectional dependence, and spatial spillovers.

The numerical results are therefore far from final. The methodological advancement in this work is more important: the joint, simultaneous estimation of the impact of two different, but often confused, phenomena: weather and climate. I defer to future research the task of refining the theoretical and empirical framework proposed here, and applying it to other macro contexts and, crucially, household and firm data.

\bibliography{weather}

\newpage
\begin{table}[htbp]\centering
\footnotesize \def\sym#1{\ifmmode^{#1}\else\(^{#1}\)\fi}
\caption{Descriptive statistics\label{tab:summ}}
\begin{tabular}{lcccccccc}
\hline\hline\\[-1.0em]
  Variable                  & Unit & Symbol &        Mean&                   Std Dev&         Min&         Max&       Obs\\\\[-1.0em]
\hline\hline\\[-1.0em]
\\[-1.0em]
Output per worker & ln(\$) & $\ln(y)$              &       9.768&              1.183&       6.047&      13.318&    7753\\
\\[-1.0em]
Capital per worker & ln(\$) & $\ln(k)$               &      10.831&              1.392&       5.650&      14.524&    7753\\
\\[-1.0em]
Temperature            & \celsius & $\bar{T}$ &      18.505&             7.269&      -1.833&      29.021&    7753\\
\\[-1.0em]
Precipitation             & cm/month & $\bar{R}$ &       9.375&           5.674&       0.299&      32.710&    7753\\
\\[-1.0em]
Standardized temperature & - &  $| z(T) |$ &     0.961&          0.737&       0.000&       7.395&    7753\\
\\[-1.0em]
Standardized precipitation  & - & $| z(R) |$ &       0.876&             0.709&       0.001&       6.717&    7753\\
Poverty dummy & - & $P$ & 0.661 & 0.473  & 0 & 1 & 7753 \\
Heat dummy & - & $H$ & 0.245 & 0.430 & 0 & 1 & 7753 \\
Polity2 & - & $G$ & 1.801 & 7.383 & -10 & 10 & 7709 \\
\hline\hline\\[-1.0em]
\end{tabular}
\end{table}

\newpage\begin{table}[htbp]\centering
\def\sym#1{\ifmmode^{#1}\else\(^{#1}\)\fi}
\caption{Baseline results.\\Dependent variable: ln(output per worker).\label{tab:base}}
\scriptsize \begin{tabular}{l*{6}{c}}
\hline\hline
                    &\multicolumn{1}{c}{(1)}&\multicolumn{1}{c}{(2)}&\multicolumn{1}{c}{(3)}&\multicolumn{1}{c}{(4)}&\multicolumn{1}{c}{(5)}&\multicolumn{1}{c}{(6)}\\
\hline
\multicolumn{7}{c}{frontier}\\
$\ln(k)$            &       0.628\sym{***}&       0.628\sym{***}&       0.634\sym{***}&       0.631\sym{***}&       0.631\sym{***}&       0.631\sym{***}\\
                    &     (79.19)         &     (79.35)         &     (77.16)         &     (77.18)         &     (76.24)         &     (77.76)         \\
[1em]
$T$                 &       0.146\sym{***}&       0.140\sym{***}&       0.147\sym{***}&       0.144\sym{***}&       0.189\sym{***}&       0.226\sym{***}\\
                    &      (6.28)         &      (3.76)         &      (3.78)         &      (4.02)         &      (5.23)         &      (8.58)         \\
[1em]
$T^2$               &    -0.00380\sym{***}&    -0.00662\sym{**} &    -0.00750\sym{**} &    -0.00730\sym{**} &    -0.00320         &    -0.00457\sym{***}\\
                    &     (-5.87)         &     (-2.75)         &     (-2.89)         &     (-3.15)         &     (-1.47)         &     (-5.42)         \\
[1em]
$R$                 &      0.0207         &      0.0328         &      0.0258         &      0.0347         &       0.246\sym{***}&       0.244\sym{***}\\
                    &      (1.77)         &      (1.34)         &      (1.01)         &      (1.42)         &      (6.75)         &      (6.89)         \\
[1em]
$R^2$               &   -0.000236         &    -0.00133         &   -0.000973         &    -0.00148         &     0.00554\sym{**} &     0.00570\sym{**} \\
                    &     (-0.69)         &     (-0.97)         &     (-0.70)         &     (-1.09)         &      (2.94)         &      (2.85)         \\
[1em]
$P \times T$        &                     &      0.0959         &      0.0867         &      0.0914         &      0.0815         &                     \\
                    &                     &      (1.73)         &      (1.57)         &      (1.66)         &      (1.41)         &                     \\
[1em]
$P \times T^2$      &                     &     0.00118         &     0.00209         &     0.00172         &    -0.00229         &                     \\
                    &                     &      (0.50)         &      (0.85)         &      (0.74)         &     (-1.03)         &                     \\
[1em]
$P \times R$        &                     &    -0.00236         &     0.00565         &    0.000946         &      -0.140\sym{**} &      -0.146\sym{**} \\
                    &                     &     (-0.08)         &      (0.19)         &      (0.03)         &     (-2.68)         &     (-3.05)         \\
[1em]
$P \times R^2$      &                     &    0.000986         &    0.000609         &    0.000988         &    -0.00599\sym{**} &    -0.00614\sym{**} \\
                    &                     &      (0.68)         &      (0.42)         &      (0.69)         &     (-3.08)         &     (-2.96)         \\
[1em]
$T \times R$        &                     &                     &                     &                     &     -0.0198\sym{***}&     -0.0199\sym{***}\\
                    &                     &                     &                     &                     &     (-8.55)         &     (-8.45)         \\
[1em]
$P \times T \times R$&                     &                     &                     &                     &      0.0167\sym{***}&      0.0172\sym{***}\\
                    &                     &                     &                     &                     &      (6.20)         &      (6.50)         \\
\hline
\multicolumn{7}{c}{inefficiency}\\
$ |z(T)| $          &      0.0456         &      0.0475         &      -0.140\sym{*}  &      -0.189\sym{**} &      -0.200\sym{***}&      -0.202\sym{***}\\
                    &      (1.27)         &      (1.30)         &     (-2.10)         &     (-3.17)         &     (-3.47)         &     (-3.54)         \\
[1em]
$ |z(R)| $          &     0.00560         &     0.00711         &      -0.174\sym{*}  &      -0.229\sym{***}&      -0.269\sym{***}&      -0.266\sym{***}\\
                    &      (0.15)         &      (0.19)         &     (-2.39)         &     (-3.86)         &     (-4.56)         &     (-4.53)         \\
[1em]
$ P \times |z(T)| $ &                     &                     &       0.241\sym{**} &       0.247\sym{***}&       0.256\sym{***}&       0.259\sym{***}\\
                    &                     &                     &      (3.24)         &      (3.70)         &      (3.95)         &      (4.05)         \\
[1em]
$ P \times |z(R)| $ &                     &                     &       0.254\sym{**} &       0.249\sym{***}&       0.283\sym{***}&       0.282\sym{***}\\
                    &                     &                     &      (3.06)         &      (3.48)         &      (4.02)         &      (4.01)         \\
[1em]
$ H \times |z(R)| $ &                     &                     &                     &       0.204\sym{**} &       0.229\sym{***}&       0.229\sym{***}\\
                    &                     &                     &                     &      (3.11)         &      (3.50)         &      (3.49)         \\
[1em]
$ H \times |z(R)| $ &                     &                     &                     &       0.201\sym{**} &       0.232\sym{**} &       0.231\sym{**} \\
                    &                     &                     &                     &      (2.83)         &      (3.26)         &      (3.25)         \\
\hline
Observations        &        7753         &        7753         &        7753         &        7753         &        7753         &        7753         \\
LpL                &      2441.1         &      2454.8         &      2487.0         &      2507.9         &      2557.9         &      2556.5         \\
\hline\hline
\multicolumn{7}{l}{\footnotesize \textit{t} statistics in parentheses}\\
\multicolumn{7}{l}{\footnotesize \sym{*} \(p<0.05\), \sym{**} \(p<0.01\), \sym{***} \(p<0.001\)}\\
\end{tabular}
\end{table}

\newpage{\scriptsize
\def\sym#1{\ifmmode^{#1}\else\(^{#1}\)\fi}
\begin{longtable}{l*{7}{c}}
\caption{Robustness checks. Dependent variable: ln(output per worker).\label{tab:robust}}\\
\hline\hline\endfirsthead\hline\endhead\hline\endfoot\endlastfoot
  &\multicolumn{1}{c}{(1)}&\multicolumn{1}{c}{(2)}&\multicolumn{1}{c}{(3)}&\multicolumn{1}{c}{(4)}&\multicolumn{1}{c}{(5)}&\multicolumn{1}{c}{(6)}&\multicolumn{1}{c}{(7)}\\
\hline
\multicolumn{8}{c}{frontier}  \\
$\ln(k)$ & 0.631\sym{***}& 0.634\sym{***}& 0.631\sym{***}& 0.628\sym{***}& 0.630\sym{***}& 0.629\sym{***}& 0.622\sym{***}\\
  & (77.76) & (79.66) & (78.92) & (78.62) & (77.05) & (79.67) & (79.73) \\
[1em]
$T$  & 0.226\sym{***}& 0.175\sym{***}& 0.225\sym{***}& 0.227\sym{***}& 0.222\sym{***}& 0.229\sym{***}& 0.278\sym{***}\\
  & (8.58) & (7.43) & (8.32) & (8.14) & (8.46) & (8.14) & (10.89) \\
[1em]
$T^2$  & -0.00457\sym{***}& -0.00337\sym{***}& -0.00446\sym{***}& -0.00449\sym{***}& -0.00451\sym{***}& -0.00452\sym{***}& -0.00666\sym{***}\\
  & (-5.42) & (-4.90) & (-5.18) & (-5.01) & (-5.30) & (-5.14) & (-8.26) \\
[1em]
$R$  & 0.244\sym{***}& 0.118\sym{***}& 0.232\sym{***}& 0.230\sym{***}& 0.253\sym{***}& 0.223\sym{***}& 0.255\sym{***}\\
  & (6.89) & (5.90) & (6.54) & (6.22) & (6.96) & (6.16) & (7.58) \\
[1em]
$R^2$  & 0.00570\sym{**} & 0.000217 & 0.00572\sym{**} & 0.00550\sym{**} & 0.00557\sym{**} & 0.00567\sym{**} & 0.00881\sym{***}\\
  & (2.85) & (0.59) & (2.83) & (2.61) & (2.78) & (2.68) & (5.44) \\
[1em]
$T \times R$ & -0.0199\sym{***}& -0.00553\sym{***}& -0.0192\sym{***}& -0.0189\sym{***}& -0.0202\sym{***}& -0.0187\sym{***}& -0.0245\sym{***}\\
  & (-8.45) & (-7.24) & (-7.92) & (-7.52) & (-8.67) & (-7.34) & (-10.70) \\
[1em]
$P \times R$ & -0.146\sym{**} & -0.148\sym{*} & -0.135\sym{**} & -0.134\sym{**} & -0.151\sym{**} & -0.130\sym{**} & -0.205\sym{***}\\
  & (-3.05) & (-2.16) & (-2.79) & (-2.69) & (-3.12) & (-2.63) & (-4.56) \\
[1em]
$P \times R^2$ & -0.00614\sym{**} &  -0.000814 & -0.00613\sym{**} & -0.00582\sym{**} & -0.00605\sym{**} & -0.00599\sym{**} & -0.00911\sym{***}\\
  & (-2.96) & (-0.94) & (-2.94) & (-2.67) & (-2.92) & (-2.75) & (-5.40) \\
[1em]
$P \times T \times R$& 0.0172\sym{***}& 0.00922\sym{***} & 0.0165\sym{***}& 0.0161\sym{***}& 0.0175\sym{***}& 0.0161\sym{***}& 0.0231\sym{***}\\
  & (6.50) & (3.31) & (6.05) & (5.67) & (6.63) & (5.67) & (9.29) \\
[1em]
$ |z(T)| $ &  &  &  &  &  & -0.000452 &  \\
  &  &  &  &  &  & (-0.13) &  \\
[1em]
$ |z(R)| $ &  &  &  &  &  & -0.000418 &  \\
  &  &  &  &  &  & (-0.12) &  \\
[1em]
$ P \times |z(T)| $ &  &  &  &  &  & -0.00469 &  \\
  &  &  &  &  &  & (-0.88) &  \\
[1em]
$ P \times |z(R)| $ &  &  &  &  &  & 0.000884 &  \\
  &  &  &  &  &  & (0.17) &  \\
[1em]
$ H \times |z(R)| $ &  &  &  &  &  & -0.00892 &  \\
  &  &  &  &  &  & (-1.35) &  \\
[1em]
$ H \times |z(R)| $ &  &  &  &  &  & 0.00177 &  \\
  &  &  &  &  &  & (0.28) &  \\
\hline
\multicolumn{8}{c}{inefficiency}  \\
$ f(z(T)) $ & -0.202\sym{***}& -0.0641 & -0.0778\sym{***} & -0.0508 &  &  &  -0.224\sym{***}\\
  & (-3.54) & (-1.52) & (-3.52) & (-1.17) &  &  & (-4.08) \\
[1em]
$ f(z(R)) $ & -0.266\sym{***}& -0.120\sym{**} & -0.0961\sym{***} & -0.0195 &  &  & -0.254\sym{***}\\
  & (-4.53) & (-2.73) & (-4.64) & (-0.48) &  &  & (-4.15) \\
[1em]
$ P \times f(z(T)) $ & 0.259\sym{***}& 0.193\sym{**} &  0.111\sym{***} & 0.0939 &  &  &  0.288\sym{***}\\
  & (4.05) & (2.96) & (4.52) & (1.79) &  &  & (4.31)\\
[1em]
$ P \times f(z(R)) $ & 0.282\sym{***}& 0.261\sym{***} & 0.0894\sym{***} & -0.0307 &  &  & 0.292\sym{***} \\
  & (4.01) & (3.65) & (3.58) & (-0.60)  &  &  & (3.72) \\
[1em]
$ H \times f(z(T)) $ & 0.229\sym{***}& 0.156\sym{*} & 0.0554\sym{*} & 0.114\sym{*} &  &  &  0.126\\
  & (3.49) & (2.47) & (2.27) & (2.00) &  &  &  (1.88)\\
[1em]
$ H \times f(z(R)) $ & 0.231\sym{**} & 0.151\sym{*} & 0.0966\sym{***} & 0.0583  &  &  & 0.168\sym{*} \\
  & (3.25) & (2.21) & (3.73) & (1.09) &  &  & (2.08) \\
[1em]
$ z(T)^+ $ &  &  &  &  & -0.148\sym{*} &  &  \\
  &  &  &  &  & (-2.25) &  &  \\
[1em]
$ z(T)^- $ &  &  &  &  & 0.327\sym{***}&  &  \\
  &  &  &  &  & (3.94) &  &  \\
[1em]
$ z(R)^+ $ &  &  &  &  & -0.306\sym{***}&  &  \\
  &  &  &  &  & (-4.62) &  &  \\
[1em]
$ z(R)^- $ &  &  &  &  & 0.258\sym{**} &  &  \\
  &  &  &  &  & (3.28) &  &  \\
[1em]
$ P \times z(T)^+ $ &  &  &  &  & 0.208\sym{**} &  &  \\
  &  &  &  &  & (2.86) &  &  \\
[1em]
$ P \times z(T)^- $ &  &  &  &  & -0.351\sym{***}&  &  \\
  &  &  &  &  & (-3.41) &  &  \\
[1em]
$ P \times z(R)^+ $ &  &  &  &  & 0.276\sym{***}&  &  \\
  &  &  &  &  & (3.37) &  &  \\
[1em]
$ P \times z(R)^- $ &  &  &  &  & -0.326\sym{***}&  &  \\
  &  &  &  &  & (-3.49) &  &  \\
[1em]
$ H \times z(T)^+ $ &  &  &  &  & 0.215\sym{**} &  &  \\
  &  &  &  &  & (3.04) &  &  \\
[1em]
$ H \times z(T)^- $ &  &  &  &  & -0.349\sym{**} &  &  \\
  &  &  &  &  & (-2.83) &  &  \\
[1em]
$ H \times z(P)^+ $ &  &  &  &  & 0.327\sym{***}&  &  \\
  &  &  &  &  & (3.57) &  &  \\
[1em]
$ H \times z(P)^- $ &  &  &  &  & -0.114 &  &  \\
  &  &  &  &  & (-1.32) &  &  \\
\hline
Observations & 7753 & 7753 & 7753 & 7753 & 7753 & 7753 & 7753 \\
LpL  & 2556.5 & 2517.9 & 2527.5 & 2499.8 & 2561.2 & 2496.2 & 2185.8 \\
\hline\hline
\multicolumn{8}{l}{\footnotesize $P$ = Poverty dummy, World Bank definition except in Column (2): 25\%ile of 1990 income distribution.}\\
\multicolumn{8}{l}{\footnotesize Columns (1), (2), (6), (7): $f(z(\cdot)) \equiv |z(\cdot)|$ ; column (3): $f(z(\cdot)) \equiv z(\cdot)^2$; column (4): $f(z(\cdot)) \equiv z(\cdot)$}\\
\multicolumn{8}{l}{\footnotesize Columns (1)-(6): Exponential distribution; column (7): Half-normal distribution.}\\
\multicolumn{8}{l}{\footnotesize \textit{t} statistics in parentheses}\\
\multicolumn{8}{l}{\footnotesize \sym{*} \(p<0.05\), \sym{**} \(p<0.01\), \sym{***} \(p<0.001\)}\\
\end{longtable}
}

\newpage{\scriptsize
\def\sym#1{\ifmmode^{#1}\else\(^{#1}\)\fi}
\begin{longtable}{l*{4}{c}}
\caption{Robustness checks. Dependent variable: ln(output per worker).\label{tab:institute}}\\
\hline\hline\endfirsthead\hline\endhead\hline\endfoot\endlastfoot
                    &\multicolumn{1}{c}{(1)}&\multicolumn{1}{c}{(2)}&\multicolumn{1}{c}{(3)}&\multicolumn{1}{c}{(4)}\\
\hline
\multicolumn{5}{c}{frontier}\\
$\ln(k)$            &       0.631\sym{***}&       0.767\sym{***}&       0.735\sym{***}&       0.664\sym{***}\\
                    &     (77.76)         &     (27.15)         &     (29.57)         &     (76.47)         \\
[1em]
$T$                 &       0.226\sym{***}&       0.175\sym{***}&       0.188\sym{***}&       0.289\sym{***}\\
                    &      (8.58)         &      (7.69)         &      (7.95)         &     (11.66)         \\
[1em]
$T^2$               &    -0.00457\sym{***}&    -0.00307\sym{***}&    -0.00342\sym{***}&    -0.00457\sym{***}\\
                    &     (-5.42)         &     (-4.52)         &     (-4.90)         &     (-6.93)         \\
[1em]
$R$                 &       0.244\sym{***}&       0.433\sym{***}&       0.382\sym{***}&       0.121\sym{***}\\
                    &      (6.89)         &      (6.73)         &      (6.43)         &      (7.25)         \\
[1em]
$R^2$               &     0.00570\sym{**} &     -0.0103\sym{***}&    -0.00950\sym{***}&    0.000712         \\
                    &      (2.85)         &     (-6.53)         &     (-6.27)         &      (1.90)         \\
[1em]
$T \times R$        &     -0.0199\sym{***}&    -0.00730\sym{***}&    -0.00630\sym{***}&    -0.00611\sym{***}\\
                    &     (-8.45)         &     (-4.57)         &     (-4.02)         &     (-9.06)         \\
[1em]
$P \times R$        &      -0.146\sym{**} &                     &                     &                     \\
                    &     (-3.05)         &                     &                     &                     \\
[1em]
$P \times R^2$      &    -0.00614\sym{**} &                     &                     &                     \\
                    &     (-2.96)         &                     &                     &                     \\
[1em]
$P \times T \times R$&      0.0172\sym{***}&                     &                     &                     \\
                    &      (6.50)         &                     &                     &                     \\
[1em]
$\ln(k)$ $\times$ $R$&                     &     -0.0277\sym{***}&     -0.0236\sym{***}&                     \\
                    &                     &     (-5.26)         &     (-4.92)         &                     \\
[1em]
$\ln(k)$ $\times$ $R^2$&                     &    0.000996\sym{***}&    0.000907\sym{***}&                     \\
                    &                     &      (6.75)         &      (6.50)         &                     \\
[1em]
$\ln(k)$ $\times$ $T \times R$&                     &    0.000108         &   0.0000654         &                     \\
                    &                     &      (1.01)         &      (0.62)         &                     \\
[1em]
Polity2             &                     &                     &                     &    -0.00350         \\
                    &                     &                     &                     &     (-1.74)         \\
[1em]
Polity2 $\times$ $R$&                     &                     &                     &   -0.000467         \\
                    &                     &                     &                     &     (-0.99)         \\
[1em]
Polity2 $\times$ $R^2$&                     &                     &                     &  0.00000887         \\
                    &                     &                     &                     &      (0.82)         \\
[1em]
Polity2 $\times$ $T \times R$&                     &                     &                     &   0.0000183\sym{*}  \\
                    &                     &                     &                     &      (2.38)         \\
\hline
\multicolumn{5}{c}{inefficiency}\\
$ |z(T)| $          &      -0.202\sym{***}&      -0.233\sym{***}&       0.849\sym{***}&      -0.201\sym{***}\\
                    &     (-3.54)         &     (-3.88)         &      (3.64)         &     (-3.61)         \\
[1em]
$ P \times |z(T)| $ &       0.259\sym{***}&       0.305\sym{***}&                     &       0.242\sym{***}\\
                    &      (4.05)         &      (4.38)         &                     &      (4.06)         \\
[1em]
$\ln(k)$ $\times$ $ |z(T)| $&                     &                     &     -0.0792\sym{***}&                     \\
                    &                     &                     &     (-3.72)         &                     \\
[1em]
$ |z(R)| $          &      -0.266\sym{***}&      -0.237\sym{***}&       0.525\sym{*}  &      -0.180\sym{***}\\
                    &     (-4.53)         &     (-3.91)         &      (1.96)         &     (-3.40)         \\
[1em]
$ P \times |z(R)| $ &       0.282\sym{***}&       0.271\sym{***}&                     &       0.178\sym{**} \\
                    &      (4.01)         &      (3.57)         &                     &      (2.87)         \\
[1em]
$\ln(k)$ $\times$ $ |z(R)| $&                     &                     &     -0.0531\sym{*}  &                     \\
                    &                     &                     &     (-2.21)         &                     \\
[1em]
$ H \times |z(T)| $ &       0.229\sym{***}&       0.182\sym{**} &       0.171\sym{**} &       0.251\sym{***}\\
                    &      (3.49)         &      (2.69)         &      (2.60)         &      (4.27)         \\
[1em]
$ H \times |z(R)| $ &       0.231\sym{**} &       0.165\sym{*}  &       0.149\sym{*}  &       0.200\sym{**} \\
                    &      (3.25)         &      (2.25)         &      (2.11)         &      (3.07)         \\
\hline
Observations        &        7753         &        7753         &        7753         &        7099         \\
LpL                &      2556.5         &      2563.6         &      2547.1         &      2621.2         \\
\hline\hline
\multicolumn{5}{l}{\footnotesize \textit{t} statistics in parentheses}\\
\multicolumn{5}{l}{\footnotesize \sym{*} \(p<0.05\), \sym{**} \(p<0.01\), \sym{***} \(p<0.001\)}\\
\end{longtable}
}

\newpage\begin{table}[htbp]\centering
\def\sym#1{\ifmmode^{#1}\else\(^{#1}\)\fi}
\caption{Cointegrating vector.\\Dependent variable: ln(output per worker)\label{tab:cv}}
\begin{tabular}{l*{5}{c}}
\hline\hline
                    &\multicolumn{1}{c}{(1)}&\multicolumn{1}{c}{(2)}&\multicolumn{1}{c}{(3)}&\multicolumn{1}{c}{(4)}&\multicolumn{1}{c}{(5)}\\
\hline
$\ln(k)$            &       0.609\sym{***}&       0.598\sym{***}&       0.599\sym{***}&       0.605\sym{***}&       0.605\sym{***}\\
                    &     (14.65)         &     (14.42)         &     (15.36)         &     (17.30)         &     (17.28)         \\
[1em]
$T$                 &                     &       0.158         &       0.124         &       0.222\sym{*}  &       0.276\sym{**} \\
                    &                     &      (1.66)         &      (1.18)         &      (2.11)         &      (2.68)         \\
[1em]
$T^2$               &                     &    -0.00697\sym{**} &     -0.0112\sym{*}  &    -0.00783         &    -0.00810\sym{***}\\
                    &                     &     (-3.26)         &     (-2.41)         &     (-1.63)         &     (-3.37)         \\
[1em]
$R$                 &                     &     -0.0234         &      0.0196         &       0.261         &       0.287\sym{*}  \\
                    &                     &     (-0.47)         &      (0.22)         &      (1.84)         &      (2.12)         \\
[1em]
$R^2$               &                     &   0.0000157         &    -0.00217         &     0.00678         &     0.00789         \\
                    &                     &      (0.01)         &     (-0.61)         &      (1.04)         &      (1.25)         \\
[1em]
$P \times T$        &                     &                     &       0.192         &       0.135         &                     \\
                    &                     &                     &      (1.00)         &      (0.66)         &                     \\
[1em]
$P \times T^2$      &                     &                     &     0.00162         &    -0.00178         &                     \\
                    &                     &                     &      (0.27)         &     (-0.28)         &                     \\
[1em]
$P \times R$        &                     &                     &     -0.0392         &      -0.222         &      -0.270         \\
                    &                     &                     &     (-0.37)         &     (-1.35)         &     (-1.78)         \\
[1em]
$P \times R^2$      &                     &                     &     0.00250         &    -0.00618         &    -0.00731         \\
                    &                     &                     &      (0.62)         &     (-0.91)         &     (-1.10)         \\
[1em]
$T \times R$        &                     &                     &                     &     -0.0235\sym{*}  &     -0.0262\sym{*}  \\
                    &                     &                     &                     &     (-1.98)         &     (-2.51)         \\
[1em]
$P \times T \times R$&                     &                     &                     &      0.0205         &      0.0242\sym{*}  \\
                    &                     &                     &                     &      (1.65)         &      (2.34)         \\
\hline
Observations        &        7753         &        7753         &        7753         &        7753         &        7753         \\
LpL                &      2186.2         &      2250.7         &      2295.4         &      2342.6         &      2339.1         \\
\hline\hline
\multicolumn{6}{l}{\footnotesize \textit{t} statistics in parentheses}\\
\multicolumn{6}{l}{\footnotesize \sym{*} \(p<0.05\), \sym{**} \(p<0.01\), \sym{***} \(p<0.001\)}\\
\end{tabular}
\end{table}

\newpage\begin{table}[htbp]\centering
\def\sym#1{\ifmmode^{#1}\else\(^{#1}\)\fi}
\caption{Short-run error-correction.\\Dependent variable: $\Delta$ln(output per worker)\label{tab:ec}}
\begin{tabular}{l*{5}{c}}
\hline\hline
                    &\multicolumn{1}{c}{(1)}&\multicolumn{1}{c}{(2)}&\multicolumn{1}{c}{(3)}&\multicolumn{1}{c}{(4)}&\multicolumn{1}{c}{(5)}\\
\hline
Output gap          &      0.0634\sym{***}&      0.0632\sym{***}&      0.0632\sym{***}&      0.0632\sym{***}&      0.0632\sym{***}\\
                    &      (7.54)         &      (7.52)         &      (7.50)         &      (7.50)         &      (7.51)         \\
[1em]
$\Delta z(T) $          &                     &    -0.00134\sym{*}  &    0.000351         &    0.000363         &    0.000408         \\
                    &                     &     (-2.46)         &      (0.65)         &      (0.66)         &      (0.79)         \\
[1em]
$\Delta z(R) $          &                     &   0.0000500         &   -0.000546         &   -0.000463         &                     \\
                    &                     &      (0.09)         &     (-0.85)         &     (-0.78)         &                     \\
[1em]
$ P \times \Delta z(T) $ &                     &                     &    -0.00253\sym{**} &    -0.00246\sym{*}  &    -0.00266\sym{**} \\
                    &                     &                     &     (-2.71)         &     (-2.29)         &     (-2.95)         \\
[1em]
$ P \times \Delta z(R) $ &                     &                     &    0.000845         &    0.000949         &                     \\
                    &                     &                     &      (0.81)         &      (0.81)         &                     \\
[1em]
$H \times \Delta z(T) $      &                     &                     &                     &   -0.000258         &                     \\
                    &                     &                     &                     &     (-0.20)         &                     \\
[1em]
$H \times \Delta z(R) $      &                     &                     &                     &   -0.000629         &                     \\
                    &                     &                     &                     &     (-0.43)         &                     \\
\hline
Observations        &        7591         &        7591         &        7591         &        7591         &        7591         \\
LpL                 &     10223.0         &     10225.8         &     10228.7         &     10228.8         &     10228.3         \\
\hline\hline
\multicolumn{6}{l}{\footnotesize \textit{t} statistics in parentheses}\\
\multicolumn{6}{l}{\footnotesize \sym{*} \(p<0.05\), \sym{**} \(p<0.01\), \sym{***} \(p<0.001\)}\\
\end{tabular}
\end{table}

\appendix
\setcounter{table}{0}
\renewcommand{\thetable}{A\arabic{table}}
\newpage 

\begin{table}[htbp]\centering
\caption{Stationarity tests}\label{tab:stat}
\begin{tabular}{lccc}
\hline\hline
series & statistic & value & p-value \\\hline
\multicolumn{4}{c}{dependent variable} \\
$ \ln(y) $ & $Z_{\bar{\tilde{t}}}$ &1.1704889 & .87909787 \\
\multicolumn{4}{c}{explanatory variables} \\
$ \ln(k) $ & $Z_{\bar{\tilde{t}}}$ &4.7080737 & .99999875 \\
$ T $ & $Z_{\bar{\tilde{t}}}$ &-5.9142383 & 1.667e-09 \\
$ T^2 $ & $Z_{\bar{\tilde{t}}}$ &-4.8816254 & 5.261e-07 \\
$ R $ & $Z_{\bar{\tilde{t}}}$ &-4.0643271 & .00002409 \\
$ R^2 $ & $Z_{\bar{\tilde{t}}}$ &-4.827023 & 6.929e-07 \\
$ T \times R $ & $Z_{\bar{\tilde{t}}}$ &-6.5881907 & 2.226e-11 \\
$ |z(T)| $ & $Z_{\bar{\tilde{t}}}$ &-47.886946 & 0 \\
$ |z(R)| $ & $Z_{\bar{\tilde{t}}}$ &-49.250722 & 0 \\
\multicolumn{4}{c}{residuals} \\
Frontier & $Z_{\bar{\tilde{t}}}$ &-.19176587 & .4239628 \\
Inefficiency & $Z_{\bar{\tilde{t}}}$ &-.42786821 & .33437354 \\
Frontier + inefficiency & $Z_{\bar{\tilde{t}}}$ &-.205783 & .41848021 \\
First differences & $Z_{\bar{\tilde{t}}}$ &1.7107859 & .95643968 \\
\hline\hline
\multicolumn{4}{l}{The null hypothesis is stationarity for each country \citep{IPS2003}.}\end{tabular}
\end{table}

{\scriptsize
\def\sym#1{\ifmmode^{#1}\else\(^{#1}\)\fi}
\begin{longtable}{l*{7}{c}}
\caption{Robustness.Dependent variable: ln(output per worker).\label{tab:trend}}\\
\hline\hline\endfirsthead\hline\endhead\hline\endfoot\endlastfoot
   &\multicolumn{1}{c}{no trend}&\multicolumn{1}{c}{linear}&\multicolumn{1}{c}{quadratic}&\multicolumn{1}{c}{cubic}&\multicolumn{1}{c}{split}&\multicolumn{1}{c}{1st diff.}&\multicolumn{1}{c}{1st diff.}\\
\hline
\multicolumn{8}{c}{frontier} \\
$\ln(k)$  & 0.692\sym{***}& 0.631\sym{***}& 0.630\sym{***}& 0.619\sym{***}& 0.622\sym{***}&   0.625\sym{***}& 0.631\sym{***}   \\
   & (117.70)  & (77.76)  & (77.91)  & (74.82)  & (75.62)  &   (21.21)  & (107.09)   \\
[1em]
$T$   & 0.254\sym{***}& 0.226\sym{***}& 0.232\sym{***}& 0.164\sym{***}& 0.218\sym{***}&   0.0909\sym{*} & 0.158\sym{***}   \\
   & (9.46)  & (8.58)  & (7.71)  & (6.11)  & (8.08)  &   (2.17)  & (11.55)   \\
[1em]
$T^2$  & -0.00244\sym{**} & -0.00457\sym{***}& -0.00463\sym{***}& -0.00440\sym{***}& -0.00392\sym{***}& -0.00233  & -0.00345\sym{***}    \\
   & (-3.10)  & (-5.42)  & (-5.31)  & (-6.01)  & (-4.65)  &   (-1.60)  & (-7.39)   \\
[1em]
$R$   & 0.233\sym{***}& 0.244\sym{***}& 0.243\sym{***}& 0.253\sym{***}& 0.258\sym{***}&  0.0408 & 0.170\sym{***}   \\
   & (6.03)  & (6.89)  & (6.82)  & (7.24)  & (7.36) & (1.78)  & (13.92) \\
[1em]
$R^2$  & 0.00516\sym{**} & 0.00570\sym{**} & 0.00574\sym{**} & 0.00459\sym{*} & 0.00685\sym{**} & 0.000405  & 0.00407\sym{***}  \\
   & (2.67)  & (2.85)  & (2.83)  & (2.45)  & (3.18)  & (0.66)  & (16.65) \\
[1em]
$T \times R$ & -0.0185\sym{***}& -0.0199\sym{***}& -0.0199\sym{***}& -0.0200\sym{***}& -0.0221\sym{***}& -0.00209  & -0.0143\sym{***} \\
   & (-6.82)  & (-8.45)  & (-8.44)  & (-8.75)  & (-9.00)  & (-1.79)  & (-20.67)  \\
[1em]
$P \times R$ & -0.138\sym{*} & -0.146\sym{**} & -0.146\sym{**} & -0.197\sym{***}& -0.212\sym{***}& -0.119\sym{*} & -0.0758\sym{***} \\
   & (-2.47)  & (-3.05)  & (-3.05)  & (-4.82)  & (-4.55)  & (-2.11)  & (-4.19)  \\
[1em]
$P \times R^2$ & -0.00438\sym{*} & -0.00614\sym{**} & -0.00617\sym{**} & -0.00461\sym{*} & -0.00664\sym{**} & -0.00121  & -0.00466\sym{***} \\
   & (-2.22)  & (-2.96)  & (-2.95)  & (-2.41)  & (-2.99)  & (-1.25)  & (-14.34)  \\
[1em]
$P \times T \times R$& 0.0135\sym{***}& 0.0172\sym{***}& 0.0172\sym{***}& 0.0183\sym{***}& 0.0202\sym{***}& 0.00594\sym{**} & 0.0119\sym{***} \\
   & (4.45)  & (6.50)  & (6.50)  & (7.42)  & (7.57)  & (2.68)  & (14.78) \\
[1em]
$\Delta \hat{u}$  &   &   &   &   &   &   & -1.605\sym{***}\\
   &   &   &   &   &   &   & (-69.35)  \\
\hline
\multicolumn{8}{c}{inefficiency} \\
$ |z(T)| $  & -0.216\sym{***}& -0.202\sym{***}& -0.203\sym{***}& -0.207\sym{***}& -0.187\sym{***}& -0.118  &   \\
   & (-3.76)  & (-3.54)  & (-3.53)  & (-3.48)  & (-3.33)  & (-1.28)  &   \\
[1em]
$ P \times |z(T)| $ & 0.257\sym{***}& 0.259\sym{***}& 0.260\sym{***}& 0.236\sym{***}& 0.228\sym{***}& 0.215  &   \\
   & (4.03)  & (4.05)  & (4.04)  & (3.46)  & (3.62)  & (1.68)  &   \\
[1em]
$ |z(R)| $  & -0.215\sym{***}& -0.266\sym{***}& -0.268\sym{***}& -0.299\sym{***}& -0.282\sym{***}& -0.106  &   \\
   & (-3.75)  & (-4.53)  & (-4.50)  & (-4.74)  & (-4.88)  & (-0.82)  &   \\
[1em]
$ P \times |z(R)| $ & 0.238\sym{***}& 0.282\sym{***}& 0.285\sym{***}& 0.318\sym{***}& 0.296\sym{***}& 0.345\sym{*} &   \\
   & (3.41)  & (4.01)  & (3.98)  & (4.13)  & (4.35)  & (2.05)  &   \\
[1em]
$ H \times |z(T)| $ & 0.189\sym{**} & 0.229\sym{***}& 0.229\sym{***}& 0.238\sym{***}& 0.246\sym{***}& 0.135  &   \\
   & (2.99)  & (3.49)  & (3.48)  & (3.50)  & (3.84)  & (1.00)  &   \\
[1em]
$ H \times |z(R)| $ & 0.183\sym{**} & 0.231\sym{**} & 0.232\sym{**} & 0.245\sym{***}& 0.259\sym{***}& -0.0492  &   \\
   & (2.70)  & (3.25)  & (3.25)  & (3.33)  & (3.73)  & (-0.28)  &   \\
[1em]
$L. |z(T)| $   &   &   &   &   &   & 0.124  &   \\
   &   &   &   &   &   & (1.38)  &   \\
[1em]
$ P \times L. |z(T)| $  &   &   &   &   &   & -0.201  &   \\
   &   &   &   &   &   & (-1.75)  &   \\
[1em]
$L. |z(R)|$   &   &   &   &   &   & -0.0305  &   \\
   &   &   &   &   &   & (-0.29)  &   \\
[1em]
$ P \times L. |z(R)| $  &   &   &   &   &   & 0.0307  &   \\
   &   &   &   &   &   & (0.91)  &   \\
[1em]
$ H \times L. |z(T)| $ &   &   &   &   &   & -0.0466  &   \\
   &   &   &   &   &   & (-0.39)  &   \\
[1em]
$ H \times L. |z(R)| $  &   &   &   &   &   & 0.0387  &   \\
   &   &   &   &   &   & (0.27)  &   \\
\hline
Observations & 7753  & 7753  & 7753  & 7753  & 7753  & 7591  & 7591  \\
LpL   & 2439.0  & 2556.5  & 2556.7  & 2775.4  & 2585.7  & 11337.1  & 19935.5  \\
\hline\hline
\multicolumn{8}{l}{\footnotesize \textit{t} statistics in parentheses}\\
\multicolumn{8}{l}{\footnotesize \sym{*} \(p<0.05\), \sym{**} \(p<0.01\), \sym{***} \(p<0.001\)}\\
\end{longtable}
}

\begin{table}[htbp]\centering
\def\sym#1{\ifmmode^{#1}\else\(^{#1}\)\fi}
\caption{Short-run error-correction.\\Dependent variable: $\Delta$ln(output per worker)\label{tab:ecabs}}
\begin{tabular}{l*{5}{c}}
\hline\hline
                    &\multicolumn{1}{c}{(1)}&\multicolumn{1}{c}{(2)}&\multicolumn{1}{c}{(3)}&\multicolumn{1}{c}{(4)}&\multicolumn{1}{c}{(5)}\\
\hline
Output gap           &      0.0634\sym{***}&      0.0634\sym{***}&      0.0634\sym{***}&      0.0634\sym{***}&      0.0633\sym{***}\\
                    &      (7.54)         &      (7.54)         &      (7.54)         &      (7.54)         &      (7.54)         \\
[1em]
$ \Delta (|z(T)|) $        &                     &    -0.00107         &     0.00120\sym{*}  &     0.00145\sym{**} &     0.00122\sym{*}  \\
                    &                     &     (-1.51)         &      (2.32)         &      (2.77)         &      (2.40)         \\
[1em]
$ \Delta (|z(R)|) $        &                     &    -0.00121         &   -0.000721         &    -0.00119         &                     \\
                    &                     &     (-1.76)         &     (-0.65)         &     (-1.18)         &                     \\
[1em]
$P \times \Delta |z(T)| $&                     &                     &    -0.00329\sym{**} &    -0.00297\sym{*}  &    -0.00338\sym{**} \\
                    &                     &                     &     (-2.92)         &     (-2.43)         &     (-3.05)         \\
[1em]
$P \times \Delta |z(R)| $&                     &                     &   -0.000632         &   -0.000951         &                     \\
                    &                     &                     &     (-0.45)         &     (-0.61)         &                     \\
[1em]
$H \times \Delta |z(T)| $&                     &                     &                     &    -0.00171         &                     \\
                    &                     &                     &                     &     (-1.07)         &                     \\
[1em]
$H \times \Delta |z(R)| $&                     &                     &                     &     0.00251         &                     \\
                    &                     &                     &                     &      (1.35)         &                     \\
\hline
Observations        &        7591         &        7591         &        7591         &        7591         &        7591         \\
LpL              &     10223.0         &     10225.3         &     10227.6         &     10229.1         &     10226.3         \\
\hline\hline
\multicolumn{6}{l}{\footnotesize \textit{t} statistics in parentheses}\\
\multicolumn{6}{l}{\footnotesize \sym{*} \(p<0.05\), \sym{**} \(p<0.01\), \sym{***} \(p<0.001\)}\\
\end{tabular}
\end{table}

\begin{table}[htbp]\centering
\caption{Stationarity tests\textemdash Error-Correction Model}\label{tab:statecm}
\begin{tabular}{cccc}
\hline\hline
model & statistic & value & p-value \\\hline
\multicolumn{4}{c}{Long-run} \\
(1) & $Z_{\bar{\tilde{t}}}$ &-.62047954 & .26747106 \\
(2) & $Z_{\bar{\tilde{t}}}$ &1.0554008 & .85437896 \\
(3) & $Z_{\bar{\tilde{t}}}$ &-2.7926792 & .00261368 \\
(4) & $Z_{\bar{\tilde{t}}}$ &-1.9118767 & .027946 \\
(5) & $Z_{\bar{\tilde{t}}}$ &-.32010646 & .37444381 \\
\multicolumn{4}{c}{Short-run} \\
(4) + (1) & $Z_{\bar{\tilde{t}}}$ &-41.963767 & 0 \\
(4) + (2) & $Z_{\bar{\tilde{t}}}$ &-42.106363 & 0 \\
(4) + (3) & $Z_{\bar{\tilde{t}}}$ &-41.939126 & 0 \\
(4) + (4) & $Z_{\bar{\tilde{t}}}$ &-41.941283 & 0 \\
(4) + (5) & $Z_{\bar{\tilde{t}}}$ &-41.95429 & 0 \\
\hline\hline
\multicolumn{4}{l}{The null hypothesis is stationarity for each country \citep{IPS2003}.}\\
\multicolumn{4}{l}{``model'' refers to columns in Tables \ref{tab:cv} and \ref{tab:ec}.}
\end{tabular}
\end{table}

\newpage 
\textbf{List of countries}\\
\\Albania\\ 
Algeria \\
Angola \\
Argentina\\ 
Armenia \\
Australia \\
Austria \\
Azerbaijan \\
Bahamas \\
Bangladesh \\
 Belarus\\ 
Belgium \\
Belize\\ 
Benin \\
Bhutan \\
Bolivia \\
Bosnia and Herzegovina\\ 
Botswana\\ 
Brazil\\ 
Brunei\\
Bulgaria\\ 
Burkina Faso \\
Burundi\\ 
Cabo Verde\\ 
Cambodia \\
                          Cameroon\\ 
                            Canada \\
          Central African Republic \\
                              Chad \\
                             Chile \\
                             China \\
                          Colombia \\
                           Comoros \\
                             Congo \\
                        Costa Rica \\
                           Croatia \\
                            Cyprus \\
                    Czech Republic \\
                 D.R. of the Congo \\
                           Denmark \\
                          Djibouti \\
                Dominican Republic \\
                           Ecuador \\
                             Egypt \\
                       El Salvador \\
                 Equatorial Guinea \\
                           Estonia \\
                          Ethiopia \\
                              Fiji \\
                           Finland \\
                            France \\
                             Gabon \\
                            Gambia \\
                           Georgia \\
                           Germany \\
                             Ghana \\
                            Greece \\
                         Guatemala \\
                            Guinea \\
                     Guinea-Bissau \\
                             Haiti \\
                          Honduras \\
                           Hungary \\
                           Iceland \\
                             India \\
                         Indonesia \\
        Iran \\
                              Iraq \\
                           Ireland \\
                            Israel \\
                             Italy \\
                             Ivory Coast\\
                           Jamaica \\
                             Japan \\
                            Jordan \\
                        Kazakhstan \\
                             Kenya \\
                            Kuwait \\
                        Kyrgyzstan \\
                   Lao People's DR \\
                            Latvia \\
                           Lebanon \\
                           Lesotho \\
                           Liberia \\
                         Lithuania \\
                        Luxembourg \\
Macedonia\\
                        Madagascar \\
                            Malawi \\
                          Malaysia \\
                              Mali \\
                        Mauritania \\
                         Mauritius \\
                            Mexico \\
                          Mongolia \\
                        Montenegro \\
                           Morocco \\
                        Mozambique \\
                           Myanmar \\
                           Namibia \\
                             Nepal \\
                       Netherlands \\
                       New Zealand \\
                         Nicaragua \\
                             Niger \\
                           Nigeria \\
                            Norway \\
                              Oman \\
                          Pakistan \\
                            Panama \\
                          Paraguay \\
                              Peru \\
                       Philippines \\
                            Poland \\
                          Portugal \\
                             Qatar \\
                 Republic of Korea \\
               Republic of Moldova \\
                           Romania \\
                Russian Federation \\
                            Rwanda \\
             Sao Tome and Principe \\
                      Saudi Arabia \\
                           Senegal \\
                            Serbia \\
                      Sierra Leone \\
                         Singapore \\
                          Slovakia \\
                          Slovenia \\
                      South Africa \\
                             Spain \\
                         Sri Lanka \\
    St. Vincent and the Grenadines \\
                    Sudan (Former) \\
                          Suriname \\
                         Swaziland \\
                            Sweden \\
                       Switzerland \\
              Syria\\
                            Taiwan \\
                        Tajikistan \\
Tanzania\\
                          Thailand \\
                              Togo \\
               Trinidad and Tobago \\
                           Tunisia \\
                            Turkey \\
                      Turkmenistan \\
                            Uganda \\
                           Ukraine \\
              United Arab Emirates \\
                    United Kingdom \\
                     United States \\
                           Uruguay \\
                        Uzbekistan \\
Venezuela \\
                          Vietnam \\
                             Yemen \\
                            Zambia \\
                          Zimbabwe\\
\\
\textbf{List of regions}\\
\\Eastern Europe and Central Asia\\
Latin America and the Caribbean\\
Middle East and North Africa\\
South and East Asia and the Pacific\\
Sub-Saharan Africa\\
Western Europe and offshoots\\
\end{document}